\documentclass[]{aa}
\usepackage[utf8]{inputenc}
\usepackage[T1]{fontenc}
\usepackage{txfonts}
\usepackage{graphicx}
\usepackage{tabularx}
\usepackage[colorlinks=true, linkcolor=blue, filecolor=blue, citecolor=blue]{hyperref}    
\usepackage{natbib}
\bibpunct{(}{)}{;}{a}{}{,} 
\makeatletter

\renewcommand*\aa@pageof{, page \thepage{} of \pageref*{LastPage}}
\makeatother

\newcommand{\orcit}[1]{\protect\href{https://orcid.org/#1}{\protect\includegraphics[width=8pt]{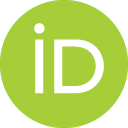}}}

\begin{document}
\title{The {\em Gaia} EDR3 view of Johnson-Kron-Cousins standard stars:\\the curated Landolt and Stetson collections\thanks{This paper is dedicated to Arlo U. Landolt, one of the fathers of photometry, who passed away on 21 January 2022.}}
\titlerunning{The Gaia view of Johnson-Kron-Cousins standards}
\author{
    E.~Pancino\orcit{0000-0003-0788-5879}\inst{\ref{oaa},\ref{ssdc}},
    P.~M.~Marrese\orcit{0000-0002-8162-3810}\inst{\ref{oarm},\ref{ssdc}},
    S.~Marinoni\orcit{0000-0001-7990-6849}\inst{\ref{oarm},\ref{ssdc}},
    N.~Sanna\orcit{0000-0001-9275-9492}\inst{\ref{oaa}},
    A.~Turchi\orcit{0000-0003-3439-8005}\inst{\ref{oaa}},
    M.~Tsantaki\orcit{0000-0002-0552-2313}\inst{\ref{oaa}},
    M.~Rainer\orcit{0000-0002-8786-2572}\inst{\ref{oaa},\ref{oami}},
    G.~Altavilla\orcit{0000-0002-9934-1352}\inst{\ref{oarm},\ref{ssdc}},
    M.~Monelli\orcit{0000-0001-5292-6380}\inst{\ref{iac},\ref{ull}},
    L.~Monaco\orcit{0000-0002-3148-9836}\inst{\ref{bello}}
    }
\authorrunning{E. Pancino et al.}

\institute{INAF -- Osservatorio Astrofisico di Arcetri, Largo E. Fermi 5, I-50125 Firenze, Italy\label{oaa}
\and Space Science Data Center -- ASI, Via del Politecnico SNC, I-00133 Rome, Italy\label{ssdc}
\and INAF -- Osservatorio Astronomico di Roma, Via Frascati 33, I-00078, Monte Porzio Catone (RM), Italy\label{oarm}
\and INAF -- Osservatorio Astronomico di Brera, via E. Bianchi 46, I-23807 Merate (LC), Italy\label{oami}
\and Instituto de Astrof\'\i sica de Canarias, Calle Via Lactea, E-38205 La Laguna, Tenerife, Spain\label{iac}
\and Facultad de F\'\i sica, Universidad de La Laguna, Avda astrofí\'\i sico Fco. Sánchez s/n, E-38200 La Laguna, Tenerife, Spain\label{ull}
\and Departamento de Ciencias Fisicas, Universidad Andres Bello, Fernandez Concha 700, Las Condes, Santiago, Chile\label{bello}
}
   
\date{Received: ---}

\abstract{In the era of large surveys and space missions, it is necessary to rely on large samples of well-characterized stars for inter-calibrating and comparing measurements from different surveys and catalogues. Among the most employed photometric systems, the Johnson-Kron-Cousins has been used for decades and for a large amount of important datasets.}{Our goal is to profit from the {\em Gaia} EDR3 data, {\em Gaia} official cross-match algorithm, and {\em Gaia}-derived literature catalogues, to provide a well-characterized and clean sample of secondary standards in the Johnson-Kron-Cousins system, as well as a set of transformations between the main photometric systems and the Johnson-Kron-Cousins one.}{Using {\em Gaia} as a reference, as well as data from reddening maps, spectroscopic surveys, and variable stars monitoring surveys, we curated and characterized the widely used Landolt and Stetson collections of more than 200\,000 secondary standards, employing classical as well as machine learning techniques. In particular, our atmospheric parameters agree significantly better with spectroscopic ones, compared to other machine learning catalogues. We also cross-matched the curated collections with the major photometric surveys to provide a comprehensive set of reliable measurements in the most widely adopted photometric systems.}{We provide a curated catalogue of secondary standards in the Johnson-Kron-Cousins system that are well-measured and as free as possible from variable and multiple sources. We characterize the collection in terms of astrophysical parameters, distance, reddening, and radial velocity. We provide a table with the magnitudes of the secondary standards in the most widely used photometric systems (ugriz, grizy, Gaia, Hipparcos, Tycho, 2MASS). We finally provide a set of 167 polynomial transformations, valid for dwarfs and giants, metal-poor and metal-rich stars, to transform $UBVRI$ magnitudes in the above photometric systems and vice-versa.}{}



\maketitle{}


\section{Introduction}

The Johnson-Kron-Cousins system is one of the most widely used photometric systems over the years. It was designed building on the work made previously by various researches, most notably on the Johnson $UBV$ \citep{johnson53,johnson63,johnson66}, Kron RI \citep{kron53} and Cousins VRI \citep{cousins76,cousins83,cousins84} photometric systems. Indeed, the Johnson system in 1966 formed the basis of most subsequent photometric systems in the optical and near infrared \citep{bessell05}. In 1992, Arlo U. Landolt published a comprehensive catalogue of equatorial standard stars which, from then on, became the fundamental defining set for the $UBVRI$ Johnson-Kron-Cousins system, and was used in the last three decades to calibrate the vast majority of all imaging observations in the $UBVRI$ passbands. The Johnson-Kron-Cousins system is mounted on optical instruments in most of the 8\,m telescopes presently available, including VLT, LBT, Subaru, or Keck.

Along the years, several photometric systems were proposed \citep[see][for a comprehensive review]{bessell05}, both based on wide bands and on medium or narrow bands that pinpoint important spectral features for specific research goals \citep[for example the Str\"omgren system:][]{stroemgren66}, while a large part of the present-day photometric surveys is based on variations of the SDSS ugriz or the Pan-STARRS grizy systems \citep[][see also Section~5 and Figure~\ref{fig:bands}]{fukugita96,tonry12}. The Johnson-Kron-Cousins system is still alive and widely employed today, but the fact that several other photometric systems are now equally or even more widely used, imposes the need to "connect" or "compare" measurements in different systems, if one wants to profitably use data from different sources. One particularly striking example is in the field of variable star studies \citep{monelli22}, where the use of long time series is one of the most fundamental tools, and the need of homogeneous photometry is of paramount importance. Photometric variability studies are in fact gradually moving from the Johnson-Kron-Cousins system \citep[for instance the OGLE or ASAS-SN surveys,][]{kaluzny95,shappee14} to ugriz or grizy systems \citep[such as the ZTF or the upcoming LSST,][]{ivezic19,chen20} and thus there is the need to accurately and precisely connect the two systems. That community has indeed started its own dedicated survey for the purpose, the AAVSO photometric all sky survey \citep[APASS\footnote{\url{https://www.aavso.org/apass}}, ][]{apass1,apass2}, which employs the B and V bands from the Johnson-Kron-Cousins system, together with the ugriz passbands from the SDSS system.

\begin{figure}[t]
    \centering
    \resizebox{\hsize}{!}{\includegraphics[clip]{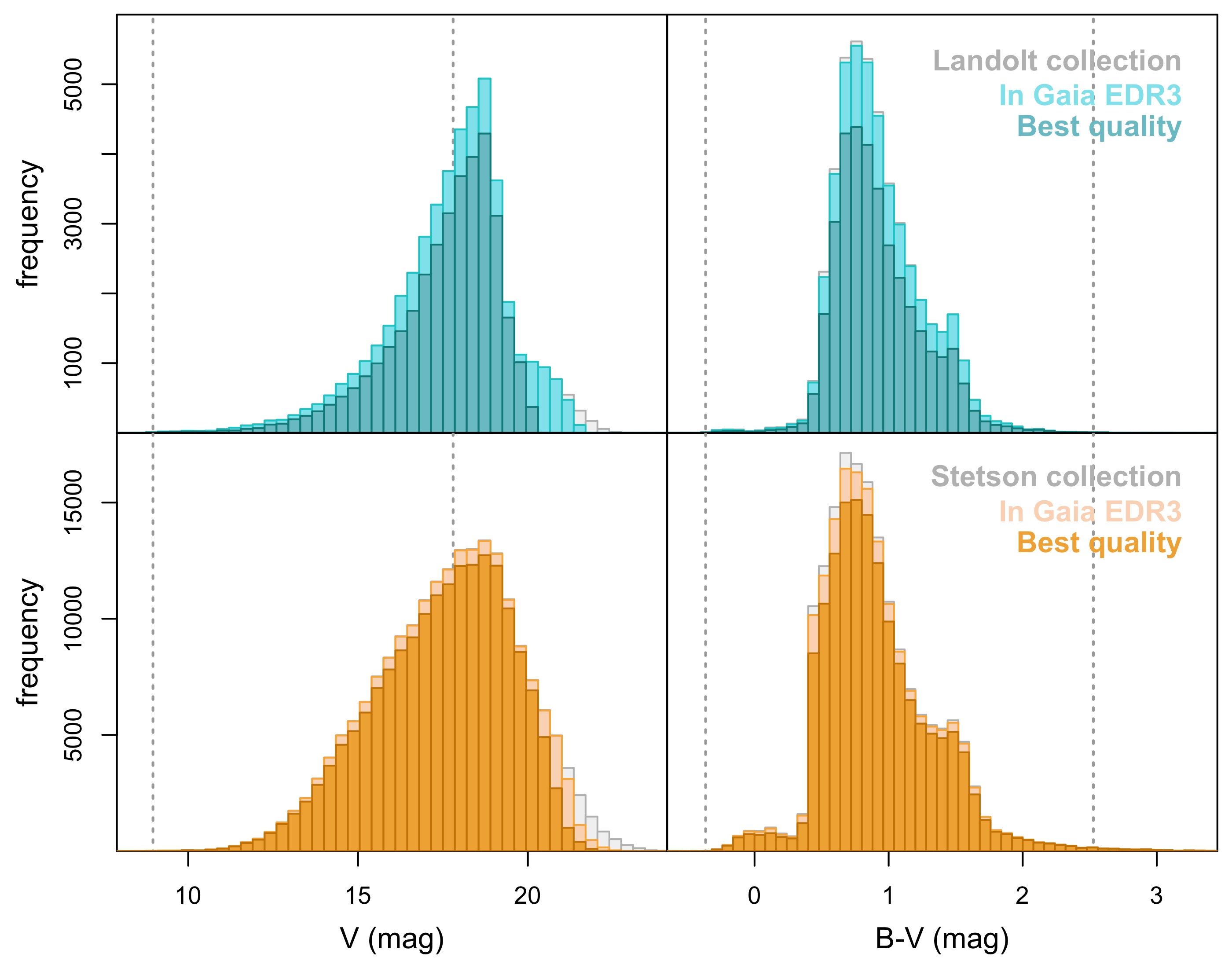}}    
    \caption{The V magnitude (left panels) and B-V color (right panels) distribution of the Landolt (top panels) and Stetson (bottom panels) collections. The full samples are plotted in grey, the samples with a match in {\em Gaia} EDR3 in light colors, and the best-quality samples (see Section~\ref{sec:gold} for details) in full colors. The dotted vertical lines show the  coverage of the original \citet{landolt92} standards set.}
    \label{fig:hist}
\end{figure}

In this paper we build on the large body of exquisite measurements by A.~U.~Landolt, \citep{landolt07a,landolt07b,landolt09,landolt13,clem13,clem16}, P.~B.~Stetson and collaborators \citep{stetson88,stetson00,stetson19}, to provide a sample of more than 200\,000 secondary standards, accurately calibrated on the original standards by \citet{landolt92}. The Landolt and Stetson collections have been curated, pruned of variable stars and binaries, and cross-matched -- with the help of {\em Gaia} -- to several large photometric surveys in different photometric systems. The two collections have also been characterized in terms of atmospheric parameters, reddening, distance, and therefore will hopefully enable a variety of calibration and inter-comparison tasks, including the calibration of Johnson-Kron-Cousins measurements from images that were not observed together with \citet{landolt92} standard fields. 

The paper is organized as follows: in Section~\ref{sec:data} we present the Landolt and the Stetson collections; in Section~\ref{sec:gaia} we describe the cross-match with {\em Gaia} and all the quality checks on the data, including the identification of binary and variable stars; in Section~\ref{sec:cat} we compare and combine the collections into one catalogue, and we characterize its stellar content in terms of distance, reddening, classification, and stellar parameters; in Section~\ref{sec:trans} we compute transformations between the Johnson-Kron-Cousins system and some other widely used photometric systems; finally, in Section~\ref{sec:concl} we summarize our results and draw our conclusions.


\section{The Landolt and Stetson collections}
\label{sec:data}


\begin{figure}[t]
    \centering
    \resizebox{\hsize}{!}{\includegraphics[clip]{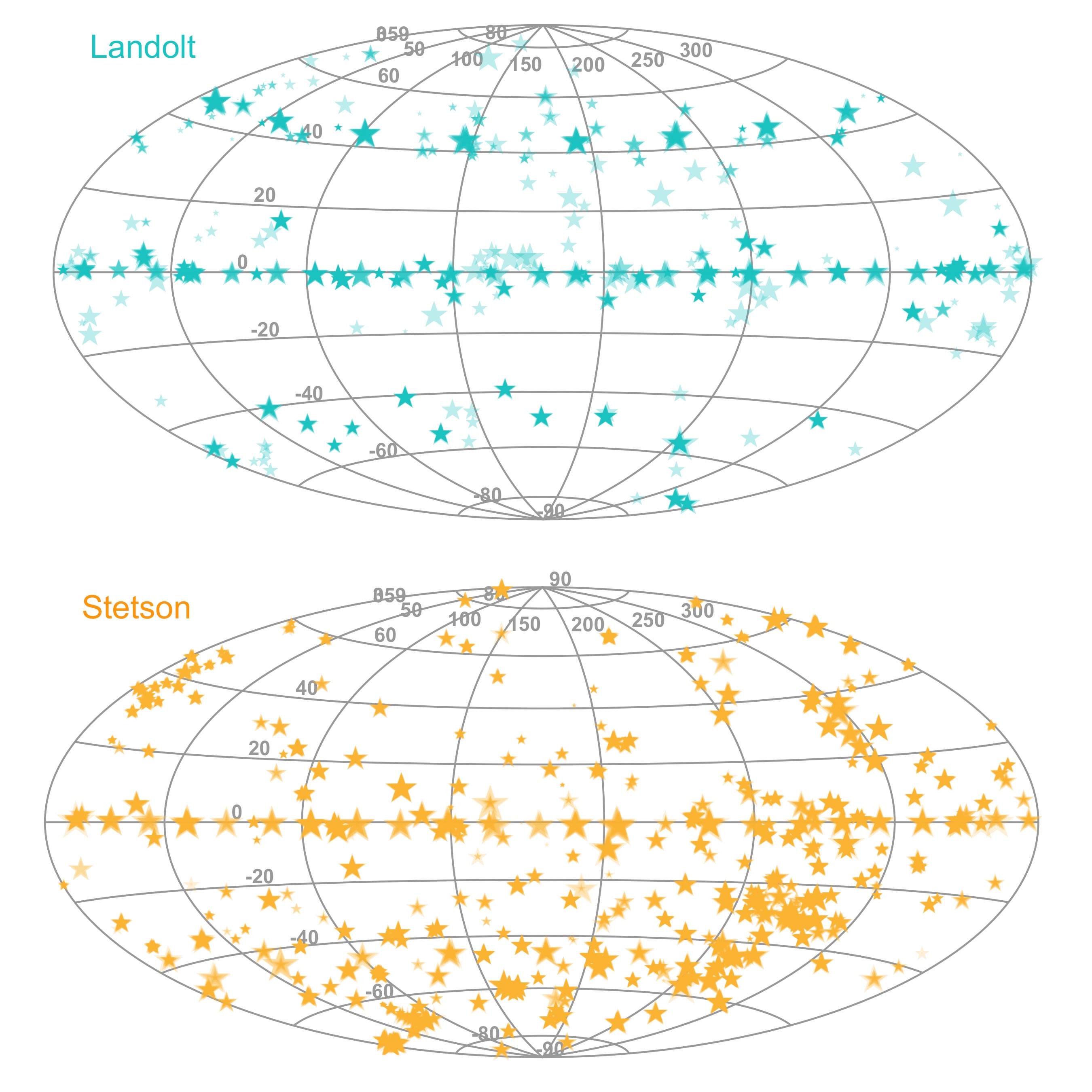}}    
    \caption{Distribution in RA and Dec on the sky of the Landolt (top panel) and Stetson (bottom panel) collections, in an Aitoff projection. Note that some of the Landolt fields are also included in the Stetson collection. The sizes of the symbols are proportional to the magnitudes and darker colors correspond to higher star densities.}
    \label{fig:map}
\end{figure}

\begin{table}
\caption{The curated Landolt collection (Section~\ref{sec:landolt}), with re-calibrated uncertainties and cross-match analysis parameters (Section~\ref{sec:lanxm}). This table contains also 520 stars with no {\em Gaia} EDR3 match (Section~\ref{sec:lanxm}), thus we report both the original Landolt coordinates and the {\em Gaia} EDR3 ICRS  coordinates (epoch J2016), when available.}
\label{tab:landolt} 
\centering                         
\begin{tabular}{lll}        
\hline\hline                
Column & Units & Description \\  
\hline                       
Unique ID    & & Unique star ID defined here \\
Gaia EDR3 ID & & Gaia Source ID from EDR3 \\ 
Star Name    & & Landolt star name \\ 
RA$_{\rm{orig}}$     & deg & Landolt original right ascension \\
Dec$_{\rm{orig}}$         & deg & Landolt original declination \\
RA$_{\rm{EDR3}}$           & deg & {\em Gaia} EDR3 right ascension \\
Dec$_{\rm{EDR3}}$          & deg & {\em Gaia} EDR3  declination \\
U            & mag & U-band magnitude \\
$\sigma$U    & mag & re-calibrated uncertainty on U \\
nU & & number of U-band measurements \\
B            & mag & B-band magnitude \\
$\sigma$B    & mag & re-calibrated uncertainty on B \\
nB & & number of B-band measurements \\
V            & mag & V-band magnitude \\
$\sigma$V    & mag & re-calibrated uncertainty on V \\
nV & & number of V-band measurements \\
R            & mag & R-band magnitude \\
$\sigma$R    & mag & re-calibrated uncertainty on R \\
nR & & number of R-band measurements \\
I            & mag & I-band magnitude \\
$\sigma$I    & mag & re-calibrated uncertainty on I \\
nI & & number of I-band measurements \\
PhotQual     & & Photometric quality flag \\
Duplicates   & & Number of duplicates ($\geq$0) \\
Neighbors    & & Number of neighbors ($\geq$1) \\
gaiaDist & arcsec & Distance from {\em Gaia} best match \\
\hline                                   \end{tabular}
\end{table}

\subsection{The Landolt collection}
\label{sec:landolt}

The Landolt standards \citep{landolt92} are the foundation of the Johnson-Kron-Cousins photometric system. Although they are tied to the Vega zero-point, they are based on the sets of primary standards in UBV by \citet{johnson63} and in RI by \citet{cousins76}. In practice, they were used for three decades to calibrate the vast majority of all imaging observations in the $UBVRI$ passbands. The original 1992 photoelectric set was later extended with observations far from the celestial equator and also with a large amount of CCD observations. We consider here a series of works by Landolt and collaborators, that can be divided into two subgroups: one based on photoelectric observations \citep{landolt92,landolt07a,landolt07b,landolt09,landolt13} and the other on CCD observations \citep{clem13,clem16}. We also complement the set with $UBVRI$ measurements for two CALSPEC\footnote{\url{https://www.stsci.edu/hst/instrumentation/reference-data-for-calibration-and-tools/astronomical-catalogs/calspec}} \citep{bohlin14,bohlin19} standards by \citet{bohlin15}. A preliminary version of the Landolt collection presented here was used to validate the Gaia Spectro-Photometric Standard Stars for the flux calibration of {\em Gaia} \citep[hereafter SPSS;][]{pancino12,pancino21,altavilla15,altavilla21,marinoni16}. A more advanced preliminary version of this collection was used to standardize the synthetic photometry obtained from the {\em Gaia} DR3 low-resolution spectra (R=$\lambda/\delta\lambda\simeq20-100$, Gaia collaboration, Montegriffo et al., in preparation).

\begin{figure*}[t]
    \centering
    \resizebox{\hsize}{!}{\includegraphics[clip]{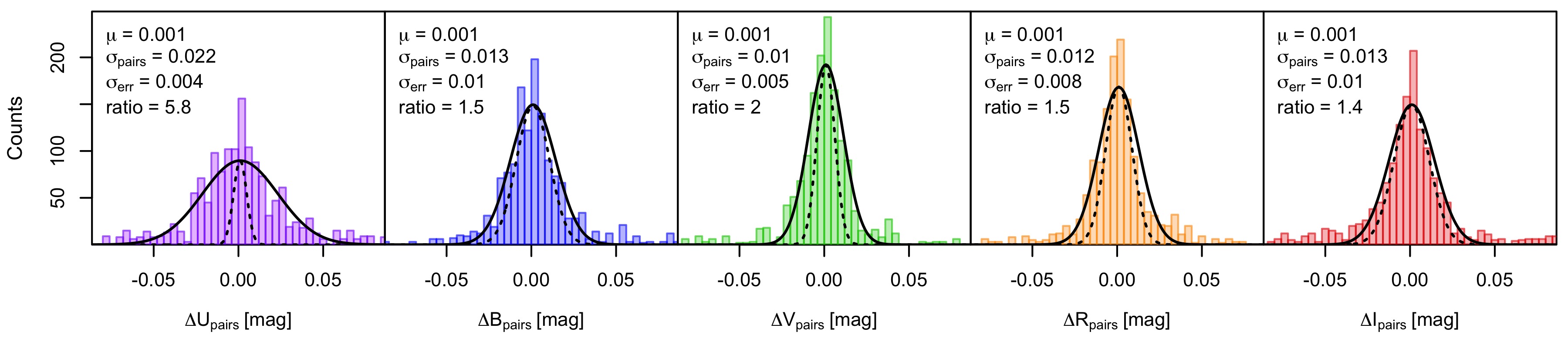}}    
    \caption{Analysis of the 1532 duplicated measurements in the Landolt collection, for 679 stars. Each panel displays the paired magnitude differences in a different band. Solid lines are Gaussians centered on the mean difference $\mu$ and standard deviation equal to the MAD (median absolute deviation) of the paired differences $\sigma_{\rm{pairs}}$. Dashed lines show Gaussians centered on $\mu$ and with a standard deviation obtained by summing in quadrature the Landolt uncertainties of the stars in each pair, $\sigma_{\rm{err}}$. The ratios between the two standard deviations are annotated in each panel, along with relevant quantities.}
    \label{fig:err}
\end{figure*}

When assembling the Landolt collection, we always chose the most recent measurement in case of stars re-observed in different papers (see also Section~\ref{sec:lanxm}). In general, newest photometry was based on larger numbers of independent measurements in each band. Detailed comparisons presented in the respective publications show that the measurements agree with the original \citet{landolt92} ones and with each other to better than 1\%, which is considered the current technological limit in our ability to calibrate the flux of stars \citep[][see also Section~\ref{sec:lanxm} and Figure~\ref{fig:err}]{clem13,bohlin19, pancino21}. Their photometry was made with a fixed aperture of 14" to replicate as much as possible the observing setup of the \citet{landolt92} set. Small trends between the newer CCD measurements and the older photoelectric measurements were carefully corrected in the \citet{clem13,clem16} analysis, to follow as faithfully as possible the photometric system defined by the original \citet{landolt92} standards. The collection contains about 47000 stars (see also Section~\ref{sec:lanxm}), of which $\simeq$90\% are from \citet{clem13,clem16}. The distribution of the Landolt collection in magnitude and color is presented in Figure~\ref{fig:hist}. The sky distribution is presented in Figure~\ref{fig:map}, where the original observations are clustered along the celestial equator, while recent observations populate the $\pm$50~deg declination strips and other areas on the sky. 

We also evaluated the relative quality of the photometry by adding to each entry the {\tt PhotQual} flag, which counts in how many passbands the star has uncertainties exceeding the 99th percentile of the uncertainty distribution as a function of magnitude. The curated Landolt collection, including indications about the number of duplicates and of neighbors from the cross-match analysis presented in Section~\ref{sec:lanxm}, is presented in Table~\ref{tab:landolt}. The uncertainties in the curated Landolt collection have been re-calibrated as described in Section~\ref{sec:lanxm}.

\subsection{The Stetson collection}
\label{sec:stetson}

The Stetson database\footnote{\url{https://www.canfar.net/storage/list/STETSON}} \citep{stetson00,stetson19} is based on more than half a million public and proprietary CCD images, collected by P.~B.~Stetson starting in 1983, when public archives did not exist yet. The images cover open and globular clusters, dwarf galaxies, supernova remnants, and other interesting areas on the sky, including some \citet{landolt92} standard fields. All images were obtained in $UBVRI$ filters and were uniformly processed. Photometry was derived with the DAOPHOT package \citep{daophot,allframe} and calibrated on the Johnson-Kron-Cousins system based on repeated \citet{landolt92} standard field observations. It is worth noting that, unlike in the Landolt collection, the Stetson collection measurements were obtained by profile fitting, and then corrected with aperture photometry curves of growth. Only stars matching strict quality criteria were considered as secondary $UBVRI$ standards:  {\em (i)} observed at least five times independently in photometric conditions, {\em (ii)} with uncertainties $<$0.02~mag, and {\em (iii)} with a spread in repeated measurements $<$0.05~mag \citep[see][for more details]{stetson19}. The Stetson database grows in time as new images are selected from public archives, and the quality of the measurements increases as more and more measurements contribute to the homogeneity and stability of the global photometric solution. Updates to the Stetson database are uploaded on a regular basis, therefore it is relevant to note that the data presented here were downloaded in April, 2021 and consists in more than 200\,000 entries (see also Section~\ref{sec:stexm}). 

The Stetson secondary standards are routinely used to calibrate photometry when no accurate standard stars observations were obtained, or to compute transformations between catalogues when no stars in common with the Landolt collection are found. For example, they were used to calibrate the photometry for the Gaia-ESO calibrating clusters observations \citep{gilmore12,pancino17}; the 2019 version of the standards was used, among other catalogues, by \citet{riello20} to validate the {\em Gaia} EDR3 photometric calibration and to derive color transformations between the {\em Gaia} and Johnson-Kron-Cousins photometry; and the transformations used to build the all-sky PLATO input catalogue (asPIC1.1) were based, among other sets, on the Stetson secondary standards as well \citep{montalto21}. 

\begin{table}
\caption{The curated Stetson collection (Section~\ref{sec:stetson}), with cross-match analysis parameters (Section~\ref{sec:stexm}), and re-calibrated uncertainties (Section~\ref{sec:comp}). The number of measurements reported here for each band refers to measurements obtained in photometric sky conditions only, according to the original Stetson data tables. This table contains also 4970 stars with no {\em Gaia} EDR3 match, thus we report both the original coordinates from Stetson and the {\em Gaia} EDR3 ICRS coordinates (epoch J2016), when available.}
\label{tab:stetson} 
\centering                         
\begin{tabular}{lll}        
\hline\hline                
Column & Units & Description \\  
\hline                       
Unique ID    & & Unique star ID defined here \\
Gaia EDR3 ID & & Gaia Source ID from EDR3 \\ 
Star Name      & & Stetson star Name \\ 
RA$_{\rm{orig}}$     & deg & Stetson original right ascension \\
Dec$_{\rm{orig}}$         & deg & Stetson original declination \\
RA$_{\rm{EDR3}}$           & deg & {\em Gaia} EDR3 right ascension \\
Dec$_{\rm{EDR3}}$          & deg & {\em Gaia} EDR3  declination \\
U            & mag & U-band magnitude \\
$\sigma$U    & mag & re-calibrated uncertainty on U \\
nU           & & Number of U-band measurements \\
B            & mag & B-band magnitude \\
$\sigma$B    & mag & re-calibrated uncertainty on B \\
nB           & & Number of B-band measurements \\
V            & mag & V-band magnitude \\
$\sigma$V    & mag & re-calibrated uncertainty on V \\
nV           & & Number of V-band measurements \\
R            & mag & R-band magnitude \\
$\sigma$R    & mag & re-calibrated uncertainty on R \\
nR           & & Number of R-band measurements \\
I            & mag & I-band magnitude \\
$\sigma$I    & mag & re-calibrated uncertainty on I \\
nI           & & Number of I-band measurements \\
VarWS        & & Welch-Stetson variability index \\
PhotQual     & & Photometric quality flag \\
Duplicates   & & Number of duplicates ($\geq$0) \\
Neighbors    & & Number of neighbors ($\geq$1) \\
gaiaDist & arcsec & Distance from {\em Gaia} best match \\
\hline                                   \end{tabular}
\end{table}

Before proceding, we used the photometric uncertainties in the Stetson collection to trim the stars with relatively worse photometric quality. We defined a quality flag ({\tt PhotQual} in Table~\ref{tab:stetson}) similarly to what done for the Landolt collection: we counted the number of bands in which the photometric error is higher than the 99th percentile, as a function of magnitude. Besides the magnitudes, errors, and the number of measurements in each band, the Stetson original catalogues contain an indication of the probability that a star is variable ({\tt VarWS} in Table~\ref{tab:stetson}), which is derived from the Welch-Stetson variability index \citep{welch93,stetson19} and the number of available measurements (see also Section~\ref{sec:stexm}). To further clean the sample, we increased the above photometric flag by one for the stars in the extreme tail of the variability flag distribution ({\tt VarWS}$>$0.04, 1588 stars). The resulting catalogue of unique secondary standards in the Stetson collection, including coordinates, photometry, quality parameters, and the cross-match analysis flags (from Section~\ref{sec:stexm}), is presented in Table~\ref{tab:stetson} (see also Figures~\ref{fig:hist} and \ref{fig:map}). The errors are re-calibrated following the analysis in Section~\ref{sec:comp}.


\section{{\em Gaia} cross-match and catalogue cleaning}
\label{sec:gaia}

We cross-matched the Landolt and Stetson collections with {\em Gaia} \citep{gaia} EDR3 data \citep{gedr3}\footnote{\url{https://www.cosmos.esa.int/web/gaia}} and with additional literature catalogues (Section~\ref{sec:cat}). As a result, we cleaned the sample from duplicates and unreliable matches and we flagged suspect binaries and variable stars, or stars with relatively lower quality in their photometry. 

We employed the {\em Gaia} cross-match software, routinely used to produce cross-identifications of {\em Gaia} sources within large public surveys\footnote{The software was developed at the SSDC and the survey cross-match results can be found in the ESA {\em Gaia} archive (\url{https://gea.esac.esa.int/archive/}) and SSDC {\em Gaia} Portal (\url{http://gaiaportal.ssdc.asi.it/}).} \citep{marrese17,marrese19}. The version of the software adopted here \citep{marrese19} finds the best matching {\em Gaia} source(s) given an external sparse catalogue, i.e., in our case the Landolt or the Stetson collection. For each source, a figure of merit is computed after correcting stellar position for the proper motions and epoch of the observations, that takes into account all the relevant uncertainties as well as the {\em Gaia} local stellar density. The software also helps in finding: {\em (i)} likely {\em duplicates}, i.e., stars in the sparse catalogue that point to the same {\em Gaia} star and {\em (ii)} {\em neighbors}, i.e., additional {\em Gaia} stars that have compatible positions with an object in the sparse catalogue. 

We also performed a cross-match between the {\em Gaia} DR2 and EDR3 releases, with the source IDs reported in Table~\ref{tab:cata}. This differs from the official {\em Gaia} DR2-EDR3 match in the sense that we looked backwards for EDR3 stars in the DR2 catalogue, while the official {\em Gaia} cross-match looks forward for DR2 stars in the EDR3 catalogue. This specific cross-match will be used below for the cross-match with the Survey of Surveys SoS \citep[][see also Sections~\ref{sec:bin} and \ref{sec:sos}]{tsantaki21} and for the reddening estimates (Section~\ref{sec:redd}).

\subsection{Landolt cross-match analysis}
\label{sec:lanxm}

For the Landolt collection, we found 563 stars without a match, of which 520 are too faint to be in {\em Gaia} (V$>$20.5~mag). For the remaining 43, which were mostly bright, high-proper motion stars, we performed manually wide cone searches (up to 5--35") and ultimately recovered them all. All of the duplicated entries found during the cross-match (679 stars with 1532 entries) came from different source papers, i.e. there were no unrecognized duplicates in any of the individual source catalogues\footnote{In several cases, the star names were slightly different (small capitals, spaces, dashes, and the like), but clearly referring to the same star.}. Among stars repeated in different papers, about a dozen pairs had large magnitude differences ($>$0.3~mag), generally in the U and B bands, but they all had very few measurements in the oldest of the source catalogues. Apart from these cases, we found that the typical spreads in the pairwise magnitude differences was about 1--2\% (Figure~\ref{fig:err}), depending on the band. The spreads of the paired differences were generally not compatible with the squared sum of the formal uncertainties on each pair, being on average larger by a varying amount. We thus multiplied the original Landolt uncertainties by the ratios reported in Figure~\ref{fig:err}, i.e., 5.8 for the U band, 1.5 for the B and R bands, 2 for the V band, and 1.4 for the I band. The re-calibrated uncertainties are reported in Table~\ref{tab:landolt}. Stars with repeated entries in different Landolt source catalogues were flagged ({\tt Duplicates} column in Table~\ref{tab:landolt}).

The resulting catalogue of unique stars contained 683 stars with one or more neighbors, i.e., with possible alternate matches, albeit with a lower figure of merit \citep[as defined by][]{marrese19}. To verify whether any neighbor could be a better match than the originally chosen best match, we computed the expected G magnitude from Landolt V and V--I, using both the relations by \citet{riello20} and the ones presented in Section~\ref{sec:edr3}, and we studied the distribution of differences between the original and expected {\em Gaia} magnitudes. On the one hand, we found that neighbors were generally fainter in {\em Gaia} than expected from the V and I magnitudes in the Landolt collection. On the other hand, the best matches had expected G magnitudes  compatible with the V and I Landolt magnitudes, within uncertanties. We thus concluded that the best matches -- according to the astrometric figure of merit -- were also best matches according to photometry. Therefore, we kept the best match in the catalogue, but we flagged all stars with neighbors ({\tt Neighbors} column in Table~\ref{tab:landolt}). 

\subsection{Stetson cross-match analysis}
\label{sec:stexm}


For the Stetson collection, the database snapshot downloaded in April 2021 contains 204\,303 individual entries. In the field of the Carina dwarf galaxy, we found almost 5000 stars with the same ID and photometry, and slightly different coordinates. These duplicates were removed before performing the cross-match and are not included in the figure above. Of the 4970 stars without a {\em Gaia} match, about 4550 are too faint to be observed by {\em Gaia}, while the remaining $\simeq$450 are mostly located in the Hydra\,I cluster field. Even when trying a wide cone search (20") and then selecting the best neighbors based on magnitudes, we could not recover unambiguously these stars in the {\em Gaia} catalogue. 

Only one unrecognized duplicate object was found in the Stetson collection\footnote{This is PG2336+004(A), with identical coordinates and different magnitudes. Only the B and V magnitudes are available, with largely discrepant values in the two entries. We adopted the weighted mean and error of the two entries.}, thus no independent statistical analysis of the reported uncertainties was possible. We therefore used the comparison with the Landolt collection to re-calibrate the Stetson uncertainties, which are reported in Table~\ref{tab:stetson} (see Section~\ref{sec:comp} for details). The stars with at least one neighbor are 19073 and the number of neighbors is reported in the {\tt Neighbors} column in Table~\ref{tab:stetson}. Similarly to the case of the Landolt collection, the neighbors do not only have a lower figure of merit based on their positions and motions, but they also tend to have {\em Gaia} magnitudes systematically fainter than expected from their V and I magnitudes, using both the relations by \citet{riello20} and our own relations from Section~\ref{sec:edr3}. Conversely, the best matches have compatible measured and expected {\em Gaia} magnitudes, within uncertainties. Therefore, we kept the best matches in all cases. 

\subsection{{\em Gaia} photometric quality flag}
\label{sec:qc}

The precision and accuracy of the {\em Gaia} EDR3 photometry is unprecedented, especially for the relatively bright stars in common with the Landolt and Stetson collections presented here: standard errors of a few mmag or better and an overall zeropoint accuracy of better than 1\% \citep[][]{riello20,pancino21}. Thanks to the numerous parameters published in the {\em Gaia} catalogue, we can perform a variety of quality assessments. In the next section we will deal with the risk of contamination and blends by nearby stars and binary companions. Here, we will use those parameters that are most relevant to evaluate the photometric measurement quality. We started by flagging (not removing) stars with: 

\begin{itemize}
    \item{a BP magnitude lower than 20.3~mag, following the recommendation by \citet{riello20};}
    \item{a two-parameter solution, i.e., without proper motion and parallax determination \citep{lindegren20}  and, what is more important here, without BP and RP magnitudes;} 
    \item{a relative error on the flux of the G, G$_{\rm{BP}}$, and G$_{\rm{RP}}$ magnitudes larger than the 99th percentile as a function of magnitude, similarly to what done in Sections~\ref{sec:landolt} and \ref{sec:stetson} for the ground-based photometric quality;}
    \item{a fraction of BP or RP contaminated transits higher than 7\% \citep[following][]{riello20}, indicating that the mean photometry could be significantly disturbed by nearby (bright) stars, located outside of the BP/RP window assigned to the star itself.}
\end{itemize}

\begin{figure}[t]
    \centering
    \resizebox{\hsize}{!}{\includegraphics[clip]{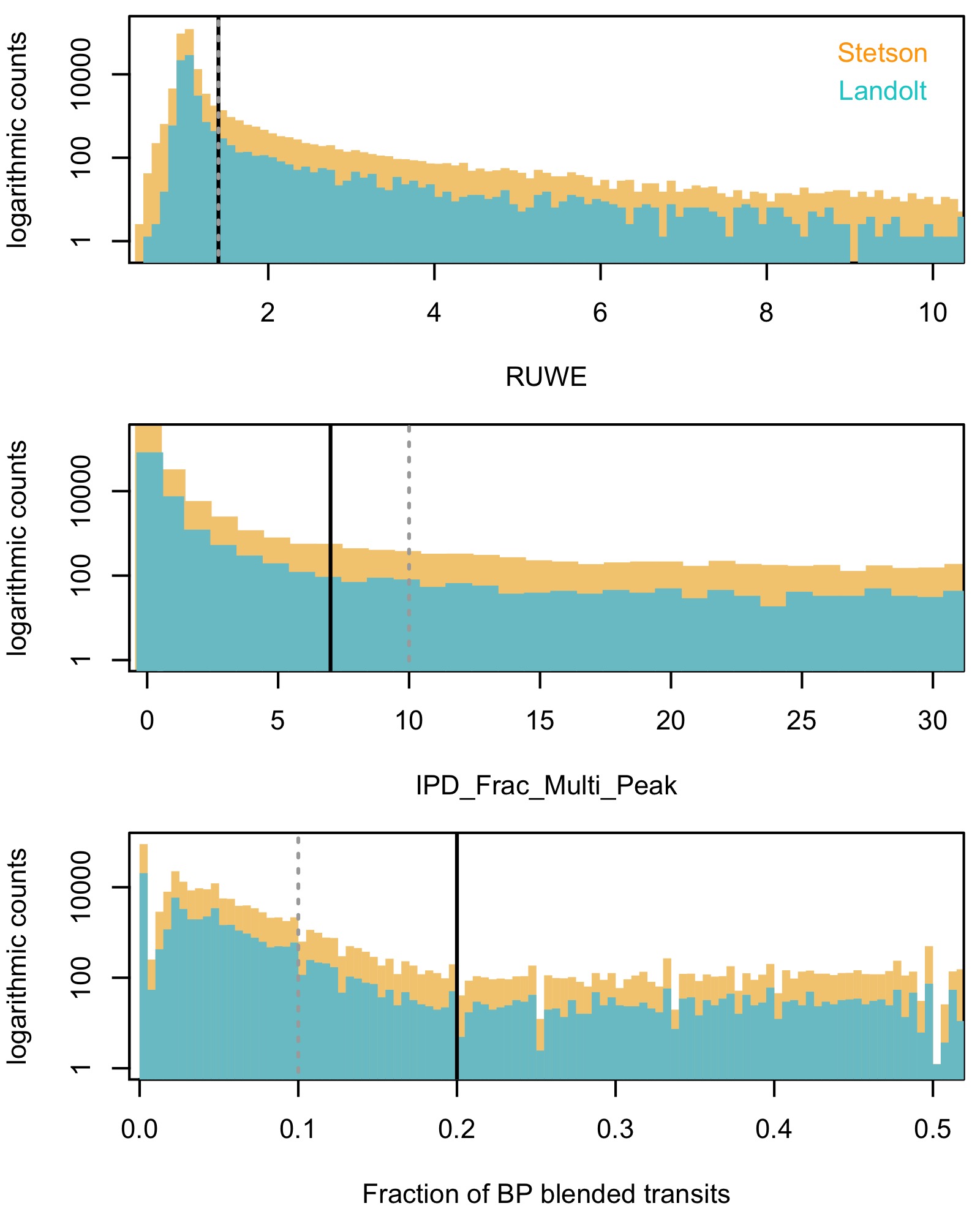}}  
    \caption{Example of quality thresholds: each panel shows the logarithmic histogram of some of the explored {\em Gaia} EDR3 parameters, for the Landolt (cyan) and Stetson (orange) collections. The recommended thresholds by \citet{lindegren20}, \citet{fabricius20}, and \citet{riello20} are indicated by grey dotted vertical lines, while our final adopted ones are indicated by black solid lines.}
    \label{fig:sele}
\end{figure}

In practice, the {\em Gaia} quality flag {\tt GaiaQual} (Table~\ref{tab:cata}) counts how many of the above quality criteria were not fulfilled, i.e., a zero flag means that all criteria were passed (best quality), one that one quality criterium was not met, and so on. In the Landolt collection, 98\% of the stars with a {\em Gaia} match met all the quality criteria, while no star failed all the criteria. In the Stetson collection, 82\% of the stars with a {\em Gaia} match passed all the quality criteria, while only one star failed all criteria. 

\subsection{Probable blends and binaries}
\label{sec:blend}

As a space observatory with a typical PSF (Point Spread Function) size of 0.177", {\em Gaia} can help in flagging potentially disturbed sources, either by chance superposition with other stars on the plane of the sky (blends) or by gravitationally bound binary companions. Among the astrometric quality indicators in the {\em Gaia} EDR3 main catalogue \citep{lindegren20,fabricius20}, we used the following for the purpose:

\begin{itemize}
\item{{\tt IPDFracOddWin} is the fraction (in percentage) of {\em Gaia} transits\footnote{We recall here that different {\em Gaia} transits are oriented with different position angles on the sky, and thus the presence of truncated windows, or the relative along-scan distance the presence of multiple peaks, can vary significantly from transit to transit. Because the angle coverage for each source changes on the sky depending on the scanning law, the best threshold can be determined empirically, depending on the specific stellar sample in hand. See also Figure~\ref{fig:sele}.} that had disturbed windows or gates, indicating that those transits are likely to be contaminated by a possible companion (or a spurious source); we conservatively flagged all stars with more than 7\% of disturbed transits;}
\item{{\tt IPDFracMultiPeak} is the fraction (in percentage) of the {\em Gaia} transits with double or multiple peaks detected in the PSF;  we conservatively flagged all stars with more than 7\% of  multiple-peak transits  \citep[Figure~\ref{fig:sele}, see also][where this indicator was used to reliably select double QSO candidates with separations of 0.3"-0.6"]{mannucci22};}
\item{{\tt RUWE} is the renormalized unit weight error of the astrometric solution; for well behaved stars it should be around one and \citet{lindegren20} suggested a cut of $<$1.4, which seems appropriate for our sample as well (Figure~\ref{fig:sele})\footnote{In the Stetson collection, some secondary standards are located in globular clusters, but they are generally relatively bright and isolated, and of good quality, thus a {\tt RUWE}$<$1.4 selection does not penalize too much the sample as far as these metal-poor stars are concerned.};}
\item{the fraction of BP and RP with blended transits, i.e. with one or more disturbing sources within the BP/RP window assigned to the star, is also a useful indicator \citep{riello20}; we flagged stars with more than 20\% of blended transits in BP or RP (Figure~\ref{fig:sele});}
\item{finally, we computed the {\tt C*} parameter, a color-corrected version of the {\tt bp\_rp\_excess\_factor} \citep{riello20}; when higher than zero, it indicates that the sum of the BP and RP fluxes is higher than the G-band flux, indicating a possible flux contamination by nearby sources; when smaller than zero, the opposite is true, for example because of background over-subtraction in BP or RP. We applied a threshold to {\tt C*} following the modeling of $\sigma_{\rm{C*}}$ by \citet[Section~9.4]{riello20}, and we flagged all sources with $|\rm{C*}|>2\,\sigma_{\rm{C*}}$.}
\end{itemize}

Similarly to what was done in the previous section, the {\tt GaiaBlend} flag (Table~\ref{tab:cata}) indicates the number of the above criteria that failed. Thus, a flag of zero means that the star passed all criteria, while a flag of $n$ means that it failed $n$ criteria. In the Landolt collection, 85\% of the stars met all criteria, while 13 stars failed them all. In the Stetson collection, 85\% of the stars met all criteria, while 39 stars failed them all. 

\begin{table}
\caption{Dictionary for the stellar classification labels in the {\tt StarType} column in Tables~\ref{tab:var} and \ref{tab:cata}. A colon ":" added after the Acronym in the {\tt StarType} column indicates an uncertain or suspected classification.}
\label{tab:dict} 
\centering 
\begin{tabular}{ll}  
\hline\hline                
Acronym & Description \\  
\hline  
BYDRA & BY Draconis type rotational variables\\
CB    & Close Binary \\
CnB   & Contact binary \\
COOL  & Cool Main Sequence star, mostly K or M \\
DCEP  & Classical Cepheid variable ($\delta$ Cephei type)\\
DSCT  & Low-amplitude $\delta$ Scuti type variables)\\
EA    & Detached Algol-type binary \\
EB    & $\beta$ Lyrae type binary\\
EW    & W Ursa Majoris type binary\\
GCAS  & $\gamma$ Cassiopeiae type variable (rapidly rotating \\
& early type stars with mass outflow)\\
HOT & Hot star, mostly OBA main sequence \\
HOT\_SD & Hot Sub-dwarf \\
L     & Red irregular variable\\
LP    & Long period variable\\
LPSR  & Long period semi-regular variable\\
MIRA  & Mira type variable\\
MS    & Main Sequence Star, mostly AFGK \\
PN    & Central star of a Planetary Nebula \\
R     & Rotational variable \\
RG    & Red giant star \\
RC    & Red clump star \\
RRAB  & RR Lyrae with asymmetric light curves, \\
      & fundamental mode \\
RRC   & RR Lyr with nearly symmetric light curves, \\
      & first overtone \\
RRD   & Double Mode RR Lyrae variables\\
ROT   & Spotted stars showing rotational modulation\\
RSCVN & RS Canum Venaticorum rotational variables\\
SB1   & Single-lined spectroscopic binary \\
SB2   & Double-lined spectroscopic binary \\
SBVC  & Spectroscopic binary or variable candidates \\
SD    & Sub-dwarf \\
SR    & Semi-regular variable\\
T2CEP & Type II Cepheid (no sub-classification)\\
T2CEP\_WVIR & W Virginis type variables (Cepheid) \\
UGER  & ER Ursae Majoris cataclysmic variables\\
UV    & Flare star \\
UWB   & Ultra-wide binary \\
V361HYA & V361 Hydrae type variable stars \\
      & (fast pulsating hot subdwarf)\\
VAR   & Variable star of unspecified type\\
WD    & White Dwarf \\
YSO   & Young Stellar Object (irregular variable) \\
ZZCET & ZZ Ceti type variable star (WD variable)\\
\hline                               
\end{tabular}
\end{table}

\subsection{Binary stars}
\label{sec:bin}

To identify spectroscopically confirmed binaries, we used the Survey of Surveys \citep[hereafter SoS,][]{tsantaki21}, which combines in a homogeneous way data from the major spectroscopic surveys (see Section~\ref{sec:sos} for more details). Among other parameters, the SoS catalogue contains a flag indicating whether a star is a spectroscopic confirmed binary in any of the surveys, using information from \citet{price20} and \citet{kounkel21} for APOGEE, \citet{traven20} for GALAH, \citet{merle17} for Gaia-ESO, \citet{birko19} for RAVE, \citet{qian19} for LAMOST, and \citet{tian20} for {\em Gaia} DR2. Because the SoS is based on {\em Gaia} DR2, we used the {\em Gaia} DR2-EDR3 cross-match described at the beginning of Section~\ref{sec:gaia}. In addition, we also searched the CoRoT \citep{deleuil18} and the Kepler \citep{kirk16} catalogues for binary stars. 

As a result, we found 301 unique confirmed binaries (308 when counting detections in multiple catalogues), of which 35 from the Landolt collection and 266 from the Stetson collection. About half of the 301 stars are classified as close binaries by APOGEE \citep{price20,kounkel21}, the others are SB2, contact binaries, radial velocity variables, or ultra-wide binaries. We have set a flag for these stars, {\tt BinFlag=1}, in Table~\ref{tab:cata}. Their classification, according to the dictionary in Table~\ref{tab:dict}, is stored in the {\tt StarType} column in Table~\ref{tab:cata}, with an annotation in the {\tt StarMethod} column stating "Binary -- see Table~\ref{tab:var}". Of the found binaries, 16 were also classified as variable stars (next section), thus we included both the binary and the variable classification in the {\tt starType} column of Table~\ref{tab:cata}, separated by a slash. More details on their classification and literature source can be found in Table~\ref{tab:var}. The spectroscopically confirmed binaries, which dominate by number the found binaries, have orbital plane inclinations that tend to be parallel to the line of sight. This makes them complementary to the suspected astrometric binaries, flagged with {\tt GaiaBlend=1} together with photometric blends, whose orbital planes tend to be perpendicular to the line of sight. 

\begin{table}
\caption{Details on notable stars with non-zero {\tt VarFlag} and/or {\tt BinFlag} in Table~\ref{tab:cata}. Stars found in more than one external catalogue have repeated entries in this table.}
\label{tab:bin} 
\label{tab:var} 
\centering                         
\begin{tabular}{ll}        
\hline\hline                
Column & Description \\  
\hline                       
Star ID    & Star ID from Table~\ref{tab:cata} \\
Gaia EDR3 ID & Gaia Source ID from EDR3 \\ 
External ID & ID in the external catalogue \\
Source & External catalogue \\
StarType & External classification (Table~\ref{tab:dict}) \\
& (":" means suspected or uncertain) \\
Notes        & Any additional notes \\
\hline                                   
\end{tabular}
\end{table}

\subsection{Variable stars}
\label{sec:var}

To search for variable stars in the Landolt and Stetson collections, we examined three different catalogues: {\em (i)} the {\em Gaia} DR2 catalogue of variable stars \citep{gdr2_var}; {\em (ii)} the ASAS-SN catalogue of variable stars \citep{shappee14,jaya18,jaya19b,jaya19a}; and {\em (iii)} the The Zwicky Transient Facility (ZTF) catalog of periodic variable stars \citep{chen20}. When appropriate, we used our {\em Gaia} DR2-EDR3 cross-match to identify the stars in the various catalogues.

We found 117 variables in the Landolt collection and 1416 in the Stetson collection (but see Section~\ref{sec:cmd} for additional identifications of young stellar objects). The majority were unclassified variables ($\simeq$900), followed by rotational variables ($\simeq$250). There were 19 variables in common between the {\em Gaia} DR2 and the ASAS-SN catalogue, 37 between ZTF and ASAS-SN, 5 between {\em Gaia} DR2 and ZTF, and only one star was reported as variable in all three catalogues. To classify variable stars present in more than one catalogue, we adopted the following choices:

\begin{itemize}
    \item{when the variability class was certain in one catalogue and uncertain in another (with a colon ":" appended), we chose the certain one (i.e., ROT over ROT:);}
    \item{if all classifications were uncertain, but one specified a class, we chose the more specific one (i.e., ROT: over VAR:);}
    \item{if two or more classifications were concordant, we chose the most specific one or the one indicating the subcategory (i.e., ROT over VAR, or RSCVN over ROT);} 
    \item{if classifications were discordant at any level (category or sub-category) we indicated them all separated by a slash (i.e., EW/EB).}
\end{itemize}

A dictionary of all the adopted {\tt StarType} labels can be found in Table~\ref{tab:dict}. In the main combined catalogue (Table~\ref{tab:cata}), we just indicated whether the star is a variable using the column {\tt VarFlag}, that is zero for non-variable stars and one for variable stars; the variable type was indicated in the {\tt StarType} column and the star was excluded from the clean sample ({\tt Qual=1}). In that table, the {\tt StarMethod} column reports the string "Variable -- see Table~\ref{tab:var}". We then provided more details in Table~\ref{tab:var}, where stars identified as variables in multiple catalogues have multiple entries, and the original classification from the corresponding literature source is reported in {\tt StarType} for each entry. 

We note here that some of the confirmed or suspected variables in the Landolt collection are part of some of the most widely used selected areas, for example around Mark\,A, Ru\,152, T\,Phe, some of the PG standards, and in the SA98, SA104, SA107, SA110, and SA113 standard fields.

\begin{figure}[t]
    \centering
    \resizebox{\hsize}{!}{\includegraphics[clip]{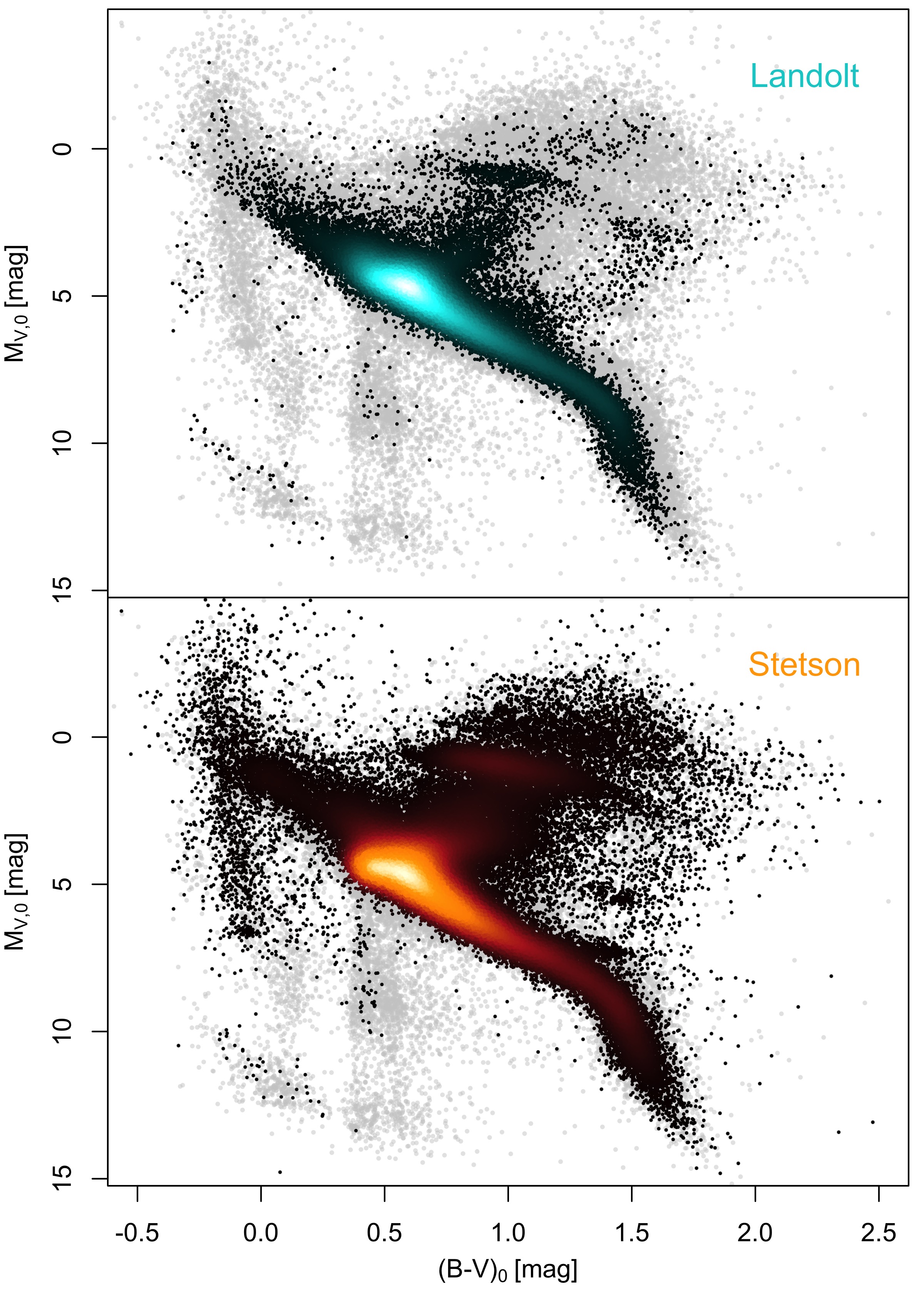}}  
    \caption{The clean Landolt (top panel) and Stetson (bottom panel) samples, in the absolute and de-reddened V, B--V plane. Stars belonging to both the Landolt and the Stetson collections are only shown in the top panel. The color scales reflect the density of points in the CMD. The whole sample of stars with distance and reddening estimates (Section~\ref{sec:redd}, Table~\ref{tab:cata}) is reported in both panels as grey dots in the background. The vertical stripes of stars between the main sequence and the white dwarfs, that almost disappear in the clean samples, are mostly made of relatively faint stars with very uncertain distance estimates.}
    \label{fig:high}
\end{figure}


\begin{table}
\caption{Content of the combined secondary standards catalogue.}
\label{tab:cata} 
\centering                         
\begin{tabular}{lll}        
\hline\hline                
Column & Units & Description \\  
\hline                       
Unique ID    & & Unique star ID defined here \\
EDR3 ID & & Gaia Source ID from EDR3 \\ 
DR2 ID & & Gaia Source ID from DR2 \\ 
Star Name    & & Star Name \\ 
Collection   & & Landolt or Stetson \\
RA$_{\rm{orig}}$     & deg & Original right ascension \\
Dec$_{\rm{orig}}$         & deg & Original declination \\
RA$_{\rm{EDR3}}$           & deg & {\em Gaia} EDR3 right ascension \\
Dec$_{\rm{EDR3}}$          & deg & {\em Gaia} EDR3  declination \\
U            & mag & U-band calibrated magnitude \\
$\sigma$U    & mag & uncertainty on U \\
nU           & & Number of U-band measurements \\
B            & mag & B-band calibrated magnitude \\
$\sigma$B    & mag & uncertainty on B \\
nB           & & Number of B-band measurements \\
V            & mag & V-band calibrated magnitude \\
$\sigma$V    & mag & uncertainty on V \\
nV           & & Number of V-band measurements \\
R            & mag & R-band calibrated magnitude \\
$\sigma$R    & mag & uncertainty on R \\
nR           & & Number of R-band measurements \\
I            & mag & I-band calibrated magnitude \\
$\sigma$I    & mag & uncertainty on I \\
nI           & & Number of I-band measurements \\
PhotQual & & Photometry quality (Section~\ref{sec:landolt}, \ref{sec:stetson}) \\
GaiaQual & & Gaia quality (Section~\ref{sec:qc}) \\
GaiaBlend & & Suspect blend or binary (Section~\ref{sec:blend}) \\
BinFlag & & Binary flag (Section~\ref{sec:bin}, Table~\ref{tab:bin}) \\
VarFlag & & Variable flag (Section~\ref{sec:var}, Table~\ref{tab:var}) \\
Qual & & Clean sample has zero (Section~\ref{sec:gold}) \\
Dist & (pc) & Distance (Section~\ref{sec:redd}) \\
Dist$_{\rm{min}}$ & (pc) & Minimum distance (Section~\ref{sec:redd}) \\
Dist$_{\rm{max}}$ & (pc) & Maximum distance (Section~\ref{sec:redd}) \\
E(B--V) & (mag) & Reddening (Section~\ref{sec:redd}) \\
$\sigma$E(B--V) & (mag) & Reddening error (Section~\ref{sec:redd}) \\
StarType & & Classification (Section~\ref{sec:cmd}, Table~\ref{tab:dict}) \\
StarMethod & & Classification method (Section~\ref{sec:cmd}) \\
RV & (km/s) & Line-of sight velocity (Section~\ref{sec:sos}) \\
$\sigma$RV & (km/s) & RV error (Section~\ref{sec:sos}) \\
RVMethod & & Source of the RV \\
T$_{\rm{eff}}$ & (K) & Effective temperature (Section~\ref{sec:sos}) \\
$\sigma$T$_{\rm{eff}}$ & (K) & T$_{\rm{eff}}$ error (Section~\ref{sec:sos}) \\
log$g$ & (dex) & Surface gravity (Section~\ref{sec:sos}) \\
$\sigma$log$g$ & (dex) & log$g$ error (Section~\ref{sec:sos}) \\
$\rm{[Fe/H]}$ & (dex) & Iron abundance (Section~\ref{sec:sos}) \\
$\sigma \rm{[Fe/H]}$ & (dex) & [Fe/H] error (Section~\ref{sec:sos}) \\
ParMethod & & Method for T$_{\rm{eff}}$, log$g$, and [Fe/H] \\ 
& & (Section~\ref{sec:sos}) \\
\hline                                   \end{tabular}
\end{table}

\section{The combined catalogue}
\label{sec:cat}

We detail in the following sections our procedures to generate the final combined catalogue: we first define a clean sample (Section~\ref{sec:gold}); we then compare the Landolt and Stetson collections before merging them (Section~\ref{sec:comp}); we further proceed to characterize the stars in terms of reddening, distance, stellar type, and astrophysical parameters (Sections~\ref{sec:redd}, \ref{sec:cmd}, \ref{sec:sos}). A summary of the catalogue content and format can be found in Table~\ref{tab:cata}, which will be available electronically and will also be published at the CDS through the Vizier\footnote{\url{https://vizier.cds.unistra.fr/}} service and at SSDC through the GaiaPortal service\footnote{\url{http://gaiaportal.ssdc.asi.it/}}.

\subsection{Clean sample}
\label{sec:gold}

Before proceeding, we build a clean sample, that will be used in the following analysis and especially in Section~\ref{sec:trans}. Note that all stars will be listed in the combined catalogue, not just the ones belonging to the clean sample. We use the criteria:

\begin{itemize}
    \item{the photometric quality flags ({\tt PhotQual}) in the Landolt (Section~\ref{sec:landolt}, Table~\ref{tab:landolt}) and Stetson collections (Section~\ref{sec:stetson}, Table~\ref{tab:stetson}) must be smaller than three, meaning that the star cannot be an outlier in three bands or more;}
    \item{the {\em Gaia} photometric quality flag {\tt GaiaQual} must also be smaller than three (Section~\ref{sec:qc});}
    \item{the {\em Gaia} suspected blends and binaries flag {\tt GaiaBlend} (Section~\ref{sec:blend}) must be zero and the star must not be a confirmed binary (i.e., {\tt BinFlag=0}) in any of the scanned binary star catalogues (Section~\ref{sec:bin});}
    \item{the star is not a suspected or confirmed variable ({\tt VarFlag} and Section~\ref{sec:var}) and in particular is not a suspected or confirmed YSO (see also Section~\ref{sec:cmd});}
    \item{as an important exception, the ten reddest stars (V--I$>$3.5~mag) in the Landolt collection with a {\em Gaia} match are kept for the following analysis, regardless of their quality flags, because of the paucity of very red stars in the combined collection.}
\end{itemize}

The clean Landolt and Stetson samples are shown in Figure~\ref{fig:high}. The Landolt clean sample contains 35194 stars (i.e., 76\%), while the Stetson one 127467 stars (i.e., 67\%), excluding those in common with the Landolt collection.  The final combined catalogue (Table~\ref{tab:cata}) contains a unified {\tt Qual} flag that is zero for the clean sample, null for stars without a {\em Gaia} match, and one for the remaining stars. 

\subsection{Comparison and catalogue merging}
\label{sec:comp}

\begin{figure}[t]
    \centering
    \resizebox{\hsize}{!}{\includegraphics[clip]{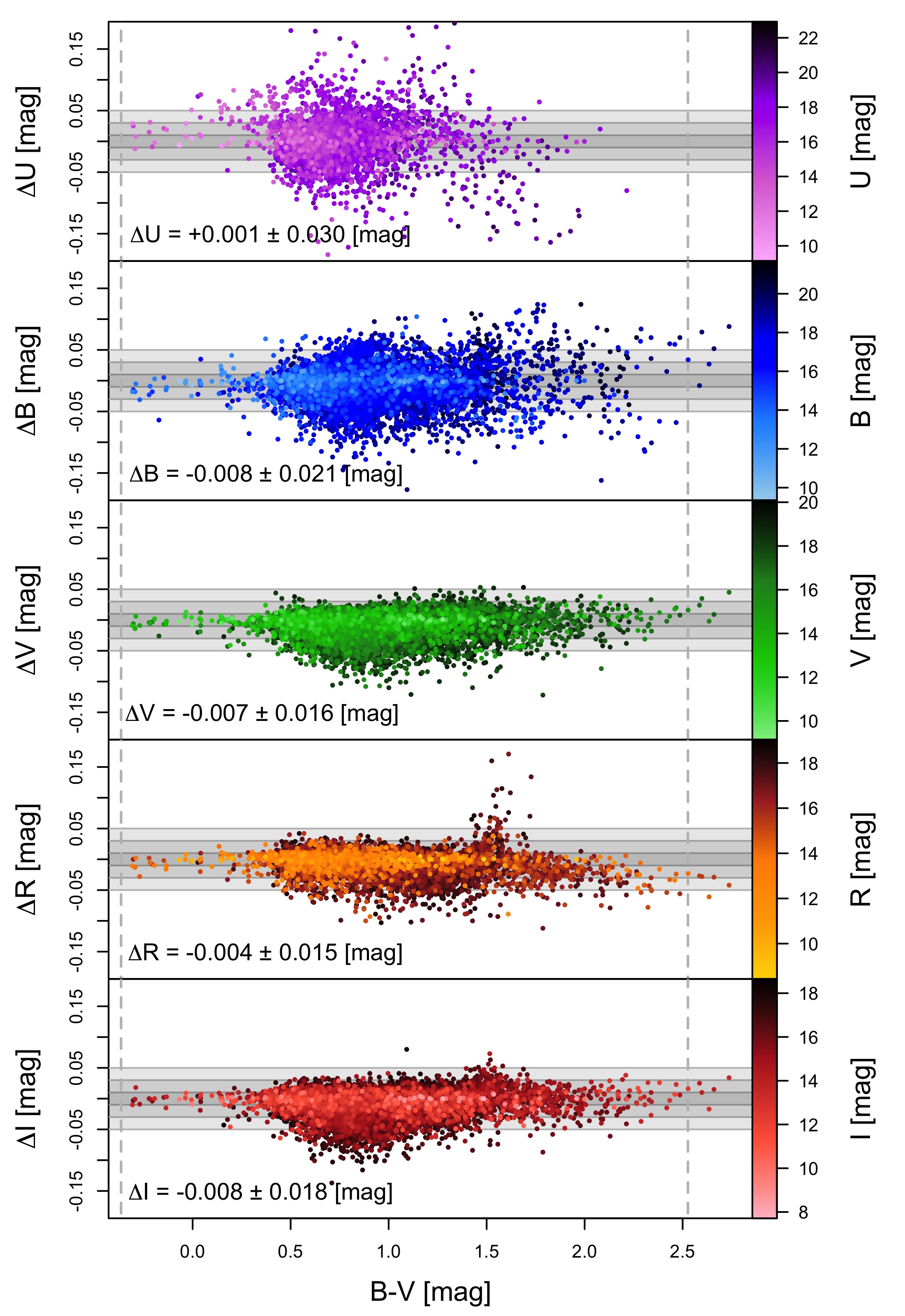}}    
    \caption{Differences between the Stetson and the Landolt magnitudes for 10769 stars in common, matching the clean-sample criteria (Section~\ref{sec:gold}). Each panel reports one of the $UBVRI$ magnitudes as a function of the Landolt B-V, as indicated. The points are colored based on the respective Landolt magnitude, and the median differences with their median absolute deviations are reported in each panel. Horizontal lines and grey-shaded areas mark the zero, $\pm$1, $\pm$3, and $\pm$5\%  differences. Vertical lines mark the color range covered by the original \citet{landolt92} catalogue, which defines the Johnson-Kron-Cousins system. The tail of faint stars extending upwards in the $\Delta$R panel at about B--V=1.5~mag is composed by M dwarfs (although other M dwarfs are well behaved) and could be populated by H${\rm{\alpha}}$ emitting stars.}
    \label{fig:comp}
\end{figure}

The Landolt and Stetson collections are accurately calibrated on the original \citet{landolt92} set of standards, which covers the color range --0.37$<$(B--V)$<$2.53 or --0.53$<$(V--I)$<$3.68~mag, with few stars in the bluest and reddest ranges. Thus, stars out of these color ranges are calibrated in extrapolation and might be less reliably placed on the standard system, and stars close to the color limits can also be uncertain, because their calibration is based on a handful of standards. In the Landolt collection, the red color limit is extended by about 0.5~mags with respect  to \citet{landolt92}, while in the Stetson collection by about one magnitude. This has to be taken into consideration when comparing the two collections with each other. For the comparison, we used stars in common between the Landolt and Stetson collections, that matched the clean-sample criteria in Section~\ref{sec:gold}: we found 10769 clean-sample stars in common between the two collections. The comparison is shown in Figure~\ref{fig:comp}. 

Given that we could not renormalize the Stetson collection uncertainties because there was only one duplicated measurement, we can use this comparison instead. If we use the mean renormalized Landolt uncertainties in each band $\sigma_{\rm{Lan}}$ (Section~\ref{sec:lanxm}) and the spread in the above comparisons $\sigma_{\rm{comp}}$, we can obtain the expected uncertainty in the Stetson collection, $\sigma_{\rm{Ste}}$ considering that
$ \sigma_{\rm{comp}}^2 = \sigma_{\rm{Lan}}^2 + \sigma_{\rm{Ste}}^2$. We thus find that the Stetson uncertainties are underestimated by approximately a factor of two. The re-calibrated uncertainties for the Landolt and Stetson collections are listed in Tables~\ref{tab:landolt} and \ref{tab:stetson}, and in the combined catalogue as well (Table~\ref{tab:cata}). 

As expected, the two collections agree to better than 1\% with each other in each band, with a spread of 1--3\% depending on the band. The U band has the highest spread of about 3\%, while the B band displays a spread of 2\%, with several discrepant stars. The bluest bands are notoriously more difficult to standardize \citep{clem13,altavilla21,pancino21} and, in addition, the Stetson collection is based on a heterogeneous collection of data from different instruments and filter sets. As anticipated, we can observe a slightly worse agreement (at the 2--3\% level) between the Landolt and Stetson collections for stars redder than B--V$\simeq$2~mag, especially in the R and I bands. The disagreement appears to be systematic, i.e., it seems to increase with color, and is most probably a consequence of the scarcity of red stars in the original \citet{landolt92} catalogue, as discussed above. After several experiments, we decided not to re-calibrate the R and I magnitudes of the Stetson or Landolt collections, because different calibrations would be required to achieve a better agreement for the M dwarfs and the M giants. However, extra caution should be used when relying on stars redder than B--V$\simeq$2.0--2.5~mag.

For the final, combined catalogue, we chose the Landolt magnitudes when available and the Stetson magnitudes when the Landolt ones were missing\footnote{We also searched for duplicates between the Landolt and Stetson stars without a {\em Gaia} EDR3 match. Only a handful were found, and the Landolt measurements were retained in the combined catalogue.}. All the stars are listed in the combined catalogue, including those lacking a {\em Gaia} counterpart and those not matching the clean-sample criteria described in Section~\ref{sec:gold}.

\subsection{Distance and reddening}
\label{sec:dist}
\label{sec:redd}

We complemented the combined catalogue with distance and reddening estimates. For Distances, we used the catalogue by \citet{bailer21}, based on the {\em Gaia} EDR3 parallaxes, who carefully accounted for the relevant parallax biases \citep{lindegren20,lindegren21}. We used their photo-geometric determination, which takes into account the expected color and magnitude distribution of stars in the Milky Way, to better constrain probable distances. The typical (median) distance is 2883~pc, while only about 25\% of the stars are farther than 4883~pc. It is worth noting that the Stetson collection contains more distant stars compared to the Landolt collection, because it targets specifically globular clusters and dwarf galaxies with long exposures. We could assign a distance estimate to 96\% of the stars in the combined collection, but for distant stars the uncertainties are of course rather high, with 25\% of the stars having distance uncertainties above 40\%.

For reddening, we explored two different sources: {\em (i)} the 3D reddening map by \citet{green19}, based on {\em Gaia} DR2, Pan-STARRS~1 \citep{ps1_1,ps1_2}, and 2MASS \citep{2mass1,2mass2}; and {\em (ii)} the 3D reddening map by \citet{lallement19}, based also on {\em Gaia} DR2 and 2MASS data. We used the 3D bins in the \citet{green19} and \citet{lallement19} maps to assign a reddening estimate to each of the stars in the Landolt and Stetson collections having a distance estimate in \citet{bailer21}. For more distant stars, or stars with large distance uncertainties, which spanned more than one 3D-bin in the maps, we apdoted the weighted average of the E(B--V) estimates corresponding to the best distance estimate, the minimum distance estimate, and the maximum one. As a result, our E(B--V) estimates often have considerably larger uncertainties than the single 3D-bin values in the original reddening maps. We studied the differences between the E(B--V) values derived in this way from the  \citet{green19} and the \citet{lallement19} maps. We found that -- except for about 8\% nearby stars with tendentially blue colors and located in specific areas of the sky -- the two sets agree very well, with a median offset of 0.01$\pm$0.05~mag and a mean one of 0.02$\pm$0.14~mag. However, the \citet{green19} set does not cover the entire sky. We therefore decided to use the \citet{lallement19} maps for sake of homogeneity\footnote{We note that all the stars with an estimate from the \citet{green19} map also had an estimate from \citet{lallement19}.}. We could assign an E(B--V) estimate to 96\% of the stars in the combined catalogue, i.e., virtually all the stars with a distance estimate. It is however very important to keep in mind that a large fraction of the stars in the combined catalogue are farther than the volume covered by the 3D maps and therefore their E(B--V) might be underestimated.

\begin{figure}[t]
    \centering
    \resizebox{\hsize}{!}{\includegraphics[clip]{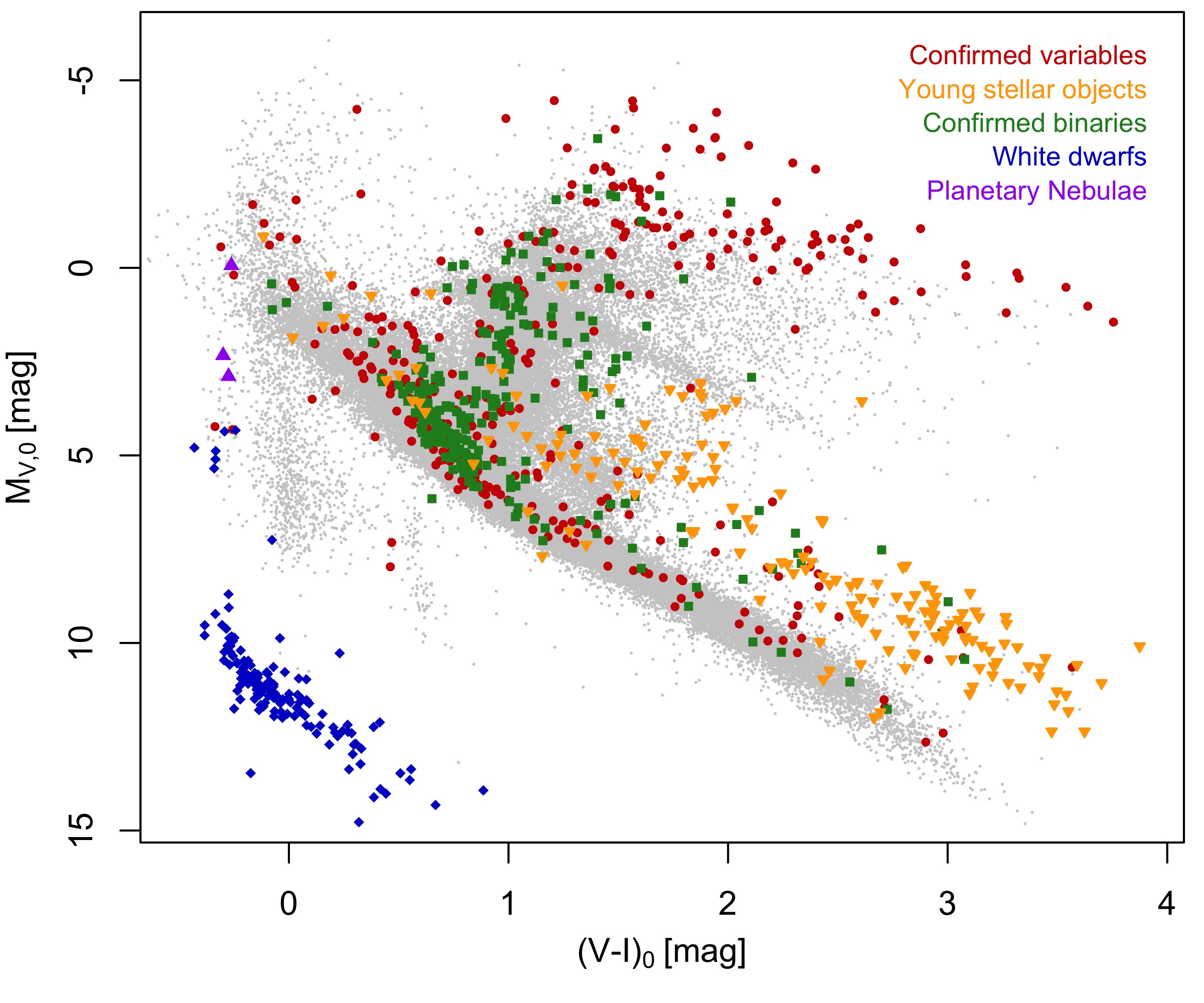}}
    \caption{The position of some stellar types in the absolute and dereddenend V, V--I plane (see Sections~\ref{sec:bin}, \ref{sec:var}, and  \ref{sec:cmd} for more details). The clean sample (Section~\ref{sec:gold}) is plotted in grey in the background.}
    \label{fig:class}
\end{figure}

\subsection{Color-magnitude diagram and stellar classification}
\label{sec:cmd}

To characterize the stellar content of the combined catalogue, we built the absolute and de-reddened color-magnitude diagram (CMD, see Figure~\ref{fig:class}), using the distance and the E(B--V) determined in Section~\ref{sec:dist}. We assumed $R_V=3.1$ and used \citet{dean78} and \citet{cardelli89} to obtain A$_{\lambda}$/A$_V$ from E(B--V). As a first step, we performed a rough manual classification of stars in different categories (main sequence, giants, red clump giants, hot, cool, subdwarfs, white dwarfs, and so on) based on their position in the above CMD, as indicated in the {\tt StarType} column in Table~\ref{tab:cata}, using the labels in Table~\ref{tab:dict}. Only stars having V and I magnitudes, as well as distance and reddening estimates, could be initially classified this way. In the same {\tt StarType} column, we also marked the following specific stellar types from literature catalogues:

\begin{itemize}
\item{five central stars of confirmed planetary nebulae,  identified with the catalogue by \citet{gonzalez21}; all five are high confidence planetary mebulae according to the authors (Group A), but only three have good $UBVRI$ photometry in the combined catalogue;}
\item{189 young stellar objects (YSO) from the \citet{marton19} catalogue; these were identified adopting slightly more stringent criteria that the ones recommended by the authors\footnote{We used R$>$0.5 and LY-$\delta$LY$\geq$0.9, or R$\leq$0.5 and SY-$\delta$SY$\geq$0.9, where R is the probability that the WISE detections are real, LY with its uncertainty $\delta$LY is the probability that the object is indeed a young stellar object in all the WISE bands, and SY with its uncertainty $\delta$SY is the same probability, but without considering the W3 and W4 bands. We classified as "YSO:", i.e., as suspect YSO, 122 additional objects matching the less-restrictive criteria recommended by the authors.}; all YSO from \citet{marton19} or from Section~\ref{sec:var}, confirmed or suspected, were flagged as variables in the {\tt VarFlag} of Table~\ref{tab:cata} and excluded from the clean sample (Section~\ref{sec:gold});}
\item{168 white dwarfs from the catalogues by \citet{kong21}, who used LAMOST spectra and literature sources to confirm their candidates, and \citet{gentile21}, who used APOGEE spectra; we used both the main and the reduced proper motion catalogues by \citet{gentile21}, applying the recommended probability cuts of 0.75\% and 0.85\%, respectively; 39 sources were in both the \citet{kong21} and \citet{gentile21} catalogues; }
\item{we also searched for stars with reliable X-ray counterparts in the optically cross-matched  XMMslew and ROSAT catalogues by \citet{salvato18} and in the eRosita catalogues by \citet{salvato21} and  \citet{robrade21}; however, we only found a handful of objects, all with non-stellar X-ray properties according to the criteria by \citet{salvato18}, that were probable false matches.} 
\end{itemize}

As an exception, for binary and variable stars, the appropriate classification from Sections~\ref{sec:bin} or \ref{sec:var} was adopted instead of the one based on the CMD. The classification is reported in the {\tt StarType} column in Table~\ref{tab:cata}, using the acronyms listed in Table~\ref{tab:dict}. The classification is accompanied by a {\tt StarMethod} column, which details whether the classification was done using the CMD, the binary or variable analysis, or one of the mentioned literature sources.

\subsection{Stellar parameters}
\label{sec:sos}

\begin{figure}[t]
    \centering
    \includegraphics[clip,width=0.98\columnwidth]{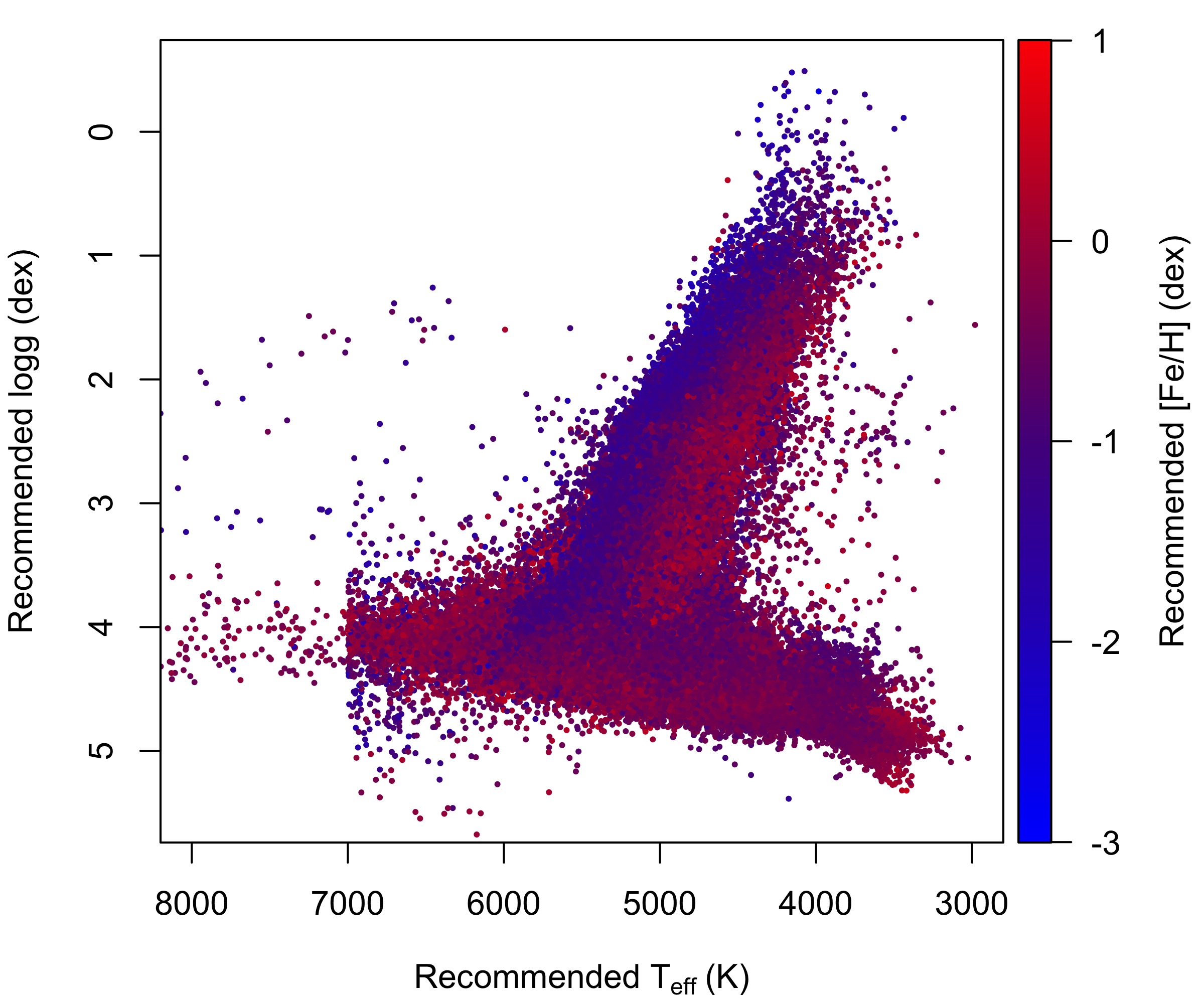}  
    \caption{The final set of recommended parameters for the clean sample. Only stars having also an [Fe/H] estimate are shown. A long tail of hot stars and the WD sequence are not shown. See Section~\ref{sec:sos} for details.} 
    \label{fig:pars}
\end{figure}

\begin{figure*}[t]
    \centering
    \includegraphics[clip,width=0.33\textwidth]{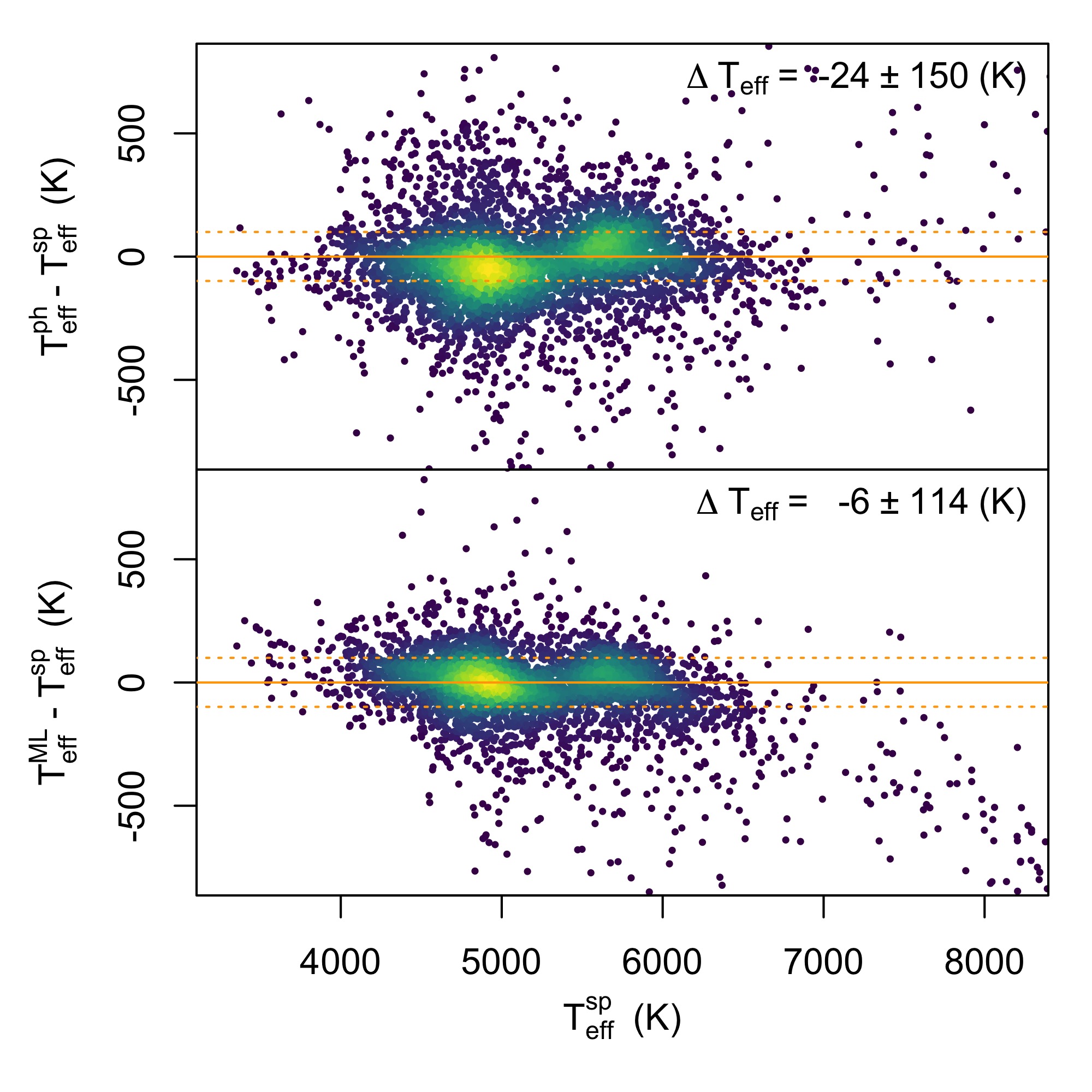}  
    \includegraphics[clip,width=0.33\textwidth]{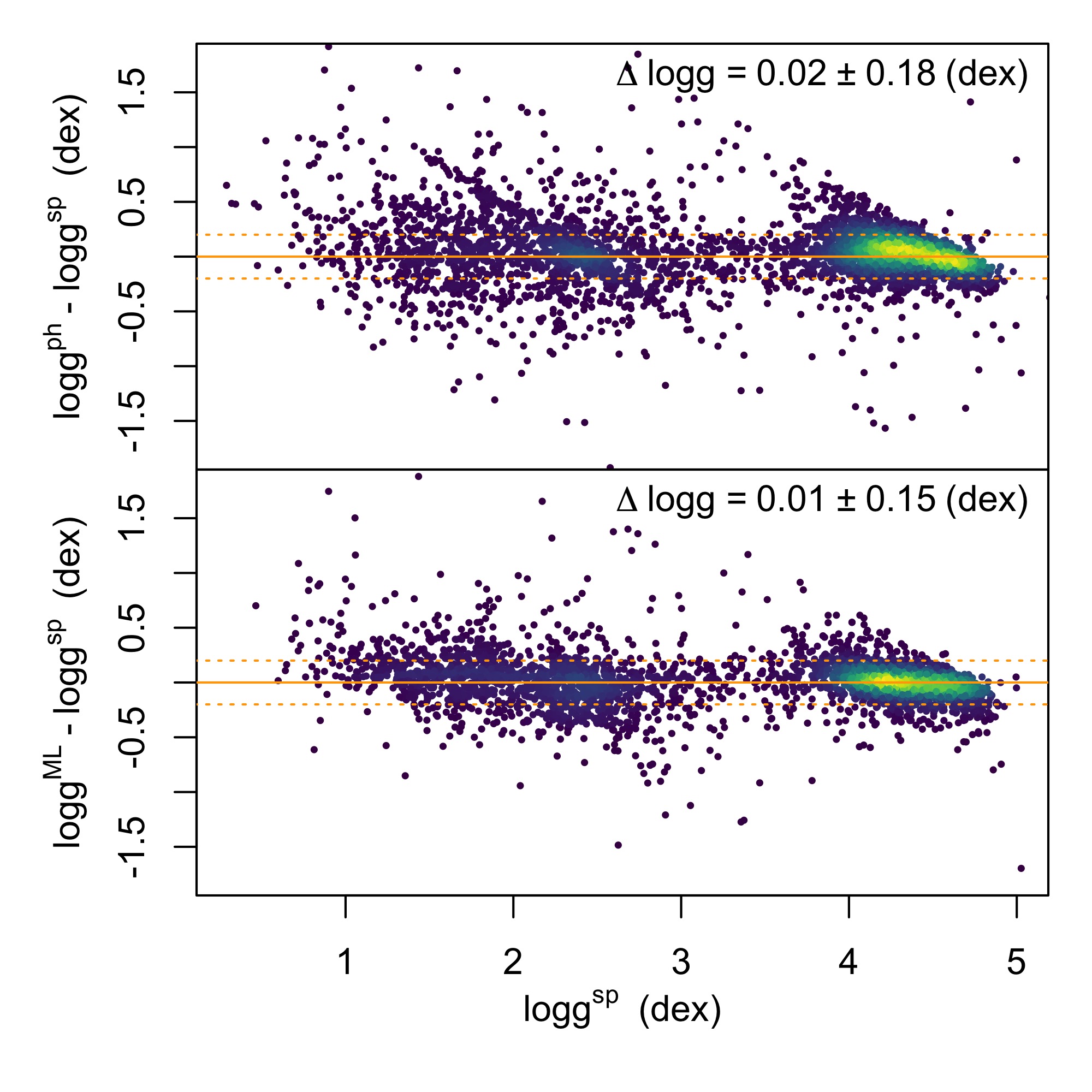} 
    \includegraphics[clip,width=0.33\textwidth]{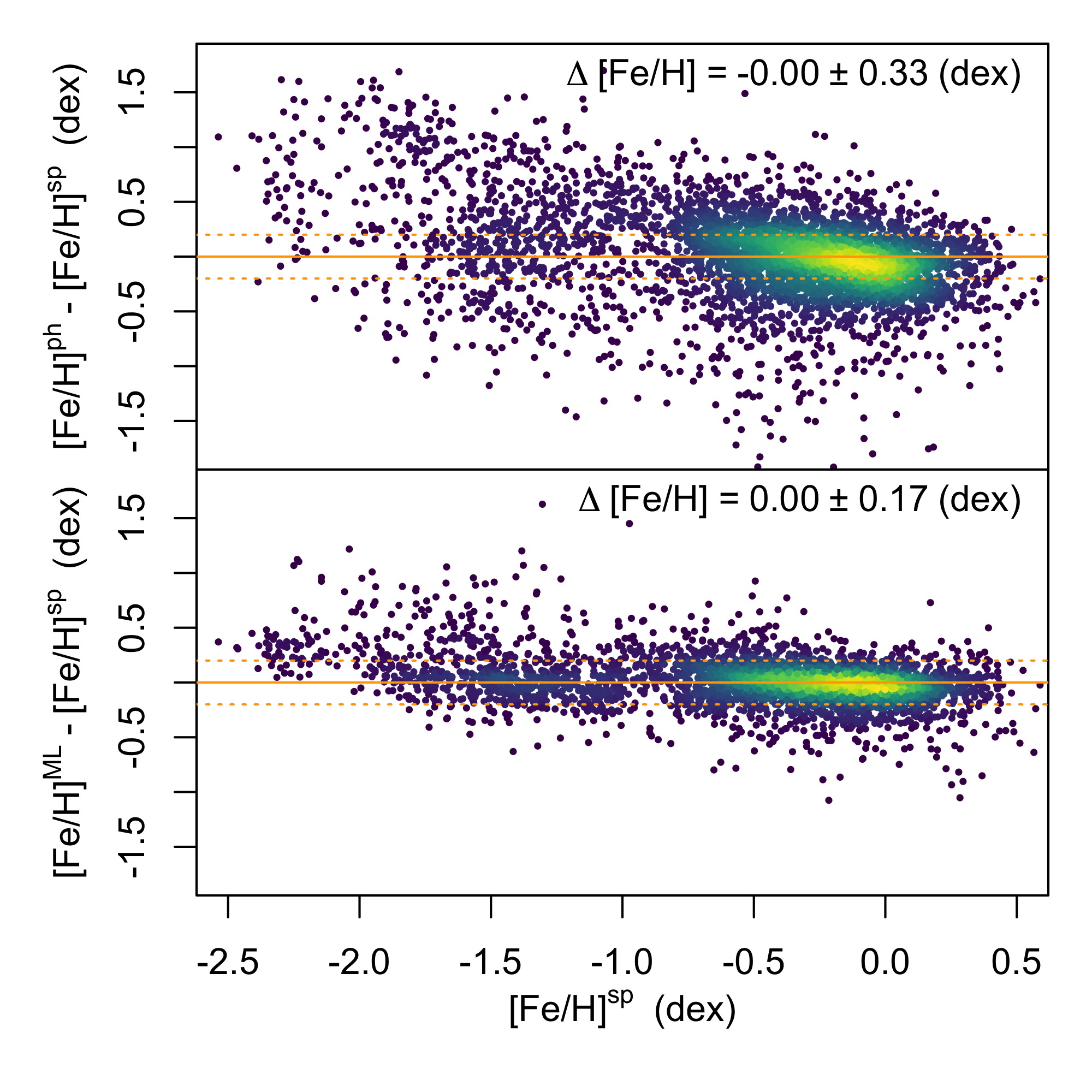}  
    \caption{Comparison between parameters obtained with different methods. The top panels show the differences between the photometric and spectroscopic parameters, the bottom ones between the machine-learning and the spectroscopic ones. The left panels show the case of T$_{\rm{eff}}$, the middle ones of log$g$, and the right ones of [Fe/H]. The median difference and MAD are written in each panel. The horizontal lines mark the zero (perfect agreement), $\pm$100~K (for T$_{\rm{eff}}$), and $\pm$0.2~dex  differences (for log$g$ and [Fe/H]), respectively. See Section~\ref{sec:sos} for more details.}
    \label{fig:parcomp}
\end{figure*}

Here we characterize the secondary standards in terms of the following parameters: line-of-sight or radial velocity, hereafter RV; effective temperature, T$_{\rm{eff}}$; surface gravity log$g$; and iron metallicity, [Fe/H]. The goal of this exercise is not to provide the most accurate parameters. Rather, we found that even a relatively good characterization of stars in the combined catalogue is sufficient to provide more reliable color transformations between photometric systems, and in particular it helps in defining the domain of applicability of those transformations (see Section~\ref{sec:trans} for more details). 
We used three different methods to derive stellar parameters: spectroscopy (Section~\ref{sec:specpars}); photometry (Section~\ref{sec:photpars}); and  machine learning (hereafter ML, Section~\ref{sec:mlpars}). We then built a recommended set of parameters by choosing the spectroscopic ones when available, then the ML ones, and for stars lacking both, we used the photometric parameters. To keep track of the method used to get parameters for each star, we added the information in Table~\ref{tab:cata}, in the {\tt parMethod} column. In this way, we could assign some estimate of the stellar parameters to 190651 stars, i.e., 80\% of the total. The final set of recommended parameters is displayed in Figure~\ref{fig:pars}, while the comparison among the results of the three methods is shown in Figure~\ref{fig:parcomp}. 

\subsubsection{Spectroscopic parameters and radial velocity}
\label{sec:specpars}

The first method we employed is based on spectroscopy. We used data from the SoS~I \citep[][see also Section~\ref{sec:bin}]{tsantaki21}, which contains homogenized, combined, and recalibrated RVs for about 11 million stars, with zero-point errors of a few hundred m/s and internal uncertainties in the range 0.5--1.5 km/s, depending on each star's survey provenance. It is based on data from {\em Gaia} DR2 \citep{gdr2}; APOGEE DR16 \citep{ahumada20}\footnote{\url{https://www.sdss.org/dr16/}}; RAVE DR6 \citep{steinmetz20}\footnote{\url{https://www.rave-survey.org/}}; GALAH DR2 \citep{buder18}\footnote{\url{https://www.galah-survey.org/}}; LAMOST DR5 \citep{zhao12}\footnote{\url{http://www.lamost.org/public/}}; and Gaia-ESO DR3 \citep{gilmore12}\footnote{\url{https://www.gaia-eso.eu/}}.

We found RV estimates for 9643 stars and stellar parameters for 6365 stars in the combined catalogue. Because only RVs are carefully re-calibrated in SoS~I, to obtain the other stellar parameters for stars observed in more than one survey, we simply took the mean of the parameters presented in Table~8 by \citet{tsantaki21}, which is  adequate for our present purpose. We complemented the SoS spectroscopic parameters using the white dwarfs identified in the \citet{kong21} or \citet{gentile21} catalogues (see also Section~\ref{sec:cmd}). This way, we could find additional RV estimates for 35 WDs and stellar parameters for 161 WDs, that were specifically derived with the use of WD synthetic spectra by the authors, comparing them with LAMOST and APOGEE spectra, respectively. For the stars in common between the two studies, we simply took the average of their values, that compared well with each other. 

\subsubsection{Photometric parameters}
\label{sec:photpars}

As a second method, we used the accurate Landolt and Stetson photometry, together with our preliminary CMD classification, to compute photometric T$_{\rm{eff}}$ and log$g$ estimates. For the T$_{\rm{eff}}$ computation we used: 

\begin{itemize}
\item{for the FGK dwarfs and giants, we used (V-I)$_0$ which is the least sensitive to metallicity, and the relations by \citet{gonzalez09} for dwarfs and giants;}
\item{for stars resulting in T$_{\rm{eff}} < 4000$~K, we still used (V-I)$_0$, but we relied on the relation by \citet{mann15}, which are computed specifically for cool stars;}
\item{for stars hotter than 7000~K and 13000~K, we used the two relations by \citet{deng20}, which however require (B--V)$_0$ rather than (V--I)$_0$;}
\item{for white dwarfs, we fitted with a third-order polynomial the (B--V)$_0$ of the WDs having spectroscopic parameters from \citet{kong21} or \citet{gentile21} and we estimated a typical 15\% uncertainty from the  residuals of the fit; we then applied the relation to all stars classified as WDs either spectroscopically or from the CMD (i.e., with {\tt StarType} WD or WD:, Section~\ref{sec:cmd}).}
\end{itemize}

We further assigned T$_{\rm{eff}}$ to 4244 stars, using the \citet{anders21} catalogue, described in more details below. As can be seen from Figure~\ref{fig:parcomp}, a good overall agreement is obtained between the spectroscopic and photometric temperatures, with a median difference of $\Delta$T$_{\rm{eff}}$=--24$\pm$150~K, in the sense that the photometric temperatures are slightly smaller than the spectroscopic ones. For consistency, we corrected the photometric T$_{\rm{eff}}$ for this small offset. However, above $\simeq$7000~K, there are various large trends and substructures -- mostly related to WDs and hot subdwarfs -- and the uncertainties can be substantially higher than for cooler stars. We thus obtained photometric T$_{\rm{eff}}$ for 93\% of the stars in the combined catalogue, i.e., the vast majority of the stars with distance and reddening estimates (Section~\ref{sec:redd}), albeit with high uncertainties above $\simeq$7000~K. 

To obtain log$g$ estimates, we computed bolometric corrections using the relations by \citet{alonso99}, \citet{flower96}, and \citet{mann15}, depending on the type of star and the validity range of the relations. With the bolometric luminosity, we estimated the radius from fundamental relations, and then log$g$ using the empirical logR$\sim$f(log$g$) relation by \citet{moya18}, which is based on a sample of stars hotter than about 5000~K. For cooler stars, both dwarfs and giants, the log$g$ estimate is much more uncertain than for warmer stars. The photometric log$g$ obtained in this way have a median difference of $\Delta$log$g$=--0.2$\pm$0.4~dex with the spectroscopic log$g$, and they show a lot of substructure, especially below 5000~K. However, we found that the log$g$ estimates obtained by means of machine learning by \citet{anders21} from photometry and astrometry (see below for more details) are in better agreement with the spectroscopic parameters and show a much smoother behavior across the CMD, especially for K and M main sequence stars. We thus used the \citet{anders21} log$g$ estimates whenever available, and the above photometric ones for 21381 hot stars for which the \citet{anders21} estimates are not available. The comparison of the log$g$ estimates obtained by combining the two approaches yields $\Delta$log$g$=+0.02$\pm$0.18~dex, and we thus obtained photometric log$g$ estimates for 91\% of the stars in the combined catalogue.

Finally, it is not possible to derive reliable estimates of [Fe/H] from simple relations based on Johnson-Kron-Cousins photometry. Thus, we used three different literature catalogues, in the following order or preference:

\begin{itemize}
\item{The \citet{shuai21} catalogue, which is based on LAMOST DR7 and {\em Gaia} EDR3 data, to provide photometric [Fe/H] estimates based on the stellar loci of stars with known metallicity; their [Fe/H] estimates show good agreement with the spectroscopic [Fe/H], with no significant trend and just a small bias: $\Delta$[Fe/H]=--0.07$\pm$0.35~dex; these were available for 24240 stars in our combined catalogue;}
\item{The \citet{anders21} catalogue mentioned above, which results from the application of the machine-learning code {\tt StarHorse} \citep{santiago16,queiroz18} to the astrometry and the photometry from {\em Gaia} EDR3, and to the combined photometry from 2MASS \citep{2mass1,2mass2}\footnote{\url{https://www.ipac.caltech.edu/project/2mass}}, Pan-STARSS1 \citep{ps1_1,ps1_2}\footnote{\url{https://panstarrs.stsci.edu/}}, Skymapper, and AllWISE \citep{cutri13}\footnote{\url{https://wise2.ipac.caltech.edu/docs/release/allwise/}}, matched using the same cross-matched algorithm used here \citep{marrese17,marrese19}; these data show a clear trend when compared to the spectroscopic [Fe/H] (Section~\ref{sec:specpars}) where the metal-poor end at about [Fe/H]$\simeq$--2.0~dex is overestimated by about 0.5--1.0~dex, with a large scatter, while the metal-rich one at about [Fe/H]$\simeq$0.5~dex is underestimated by about 0.5~dex; we obtained photometric [Fe/H] estimates for 126706 stars, albeit less accurate than the \citet{shuai21} ones;}
\item{the [Fe/H] estimates by \citet{miller15}, based on SDSS 10 photometry \citep{ahn14}\footnote{\url{https://www.sdss3.org/dr10/}}, which are available for 15523 stars in our combined catalogue; these estimates show a very similar trend with the spectroscopic [Fe/H] estimates as the \citet{anders21} ones, therefore, we used them for the 409 stars that did not have an estimate by \citet{anders21} or \citet{shuai21};}
\item{the [Fe/H] estimates by \citet{chiti21}, based on SkyMapper photometry \citep{wolf18}, which are available for 998 stars in our combined catalogue; these saturate at about [Fe/H]$\simeq$--1.0~dex but are linearly correlated with  the spectroscopic estimates below [Fe/H]$\simeq$--1.2~dex; we used them for 12 additional stars without an estimate by other sources.}
\end{itemize} 

The comparison between photometric and spectroscopic parameters is shown in the top panels of Figure~\ref{fig:parcomp}.

\subsubsection{Machine learning parameters}
\label{sec:mlpars}

Having a sizeable set of stars with reliable spectroscopic parameters from the SoS, and keeping in mind the difficulties of obtaining log$g$ and [Fe/H] estimates for field stars from photometry (see Figure~\ref{fig:parcomp}, top panels), we experimented with ML algorithms as well. It has indeed already been shown that purely numerical methods, such as ML techniques, can be successfully employed to estimate stellar parameters from photometric inputs \citep[see the cited work by][and references therein]{miller15,santiago16,queiroz18,shuai21,anders21}. 

There is a large number of ML techniques in the literature that can apply to the case presented in this study. The problem at hand, in the ML lexicon, can be categorized as a supervised regression problem. Supervised ML methods are trained on a relatively small sample of objects, where the expected output is known (in our case, the spectroscopic T$_{\rm{eff}}$, log$g$ or [Fe/H] from Section~\ref{sec:specpars}). Over this training sample, the algorithm learns how to transform the input variables, which can be an arbitrary number (see Table~\ref{tab:mlinput}), into the output estimates by minimizing the resulting error. The trained algorithm is then tested on an independent sample, where the output is still known but is not provided to the algorithm, in order to prove that it can indeed work in a general case. The size of the sample used for both training and testing is crucial to maximize the performances of the method. We tested several ML methods, namely: Random Forest \citep[RF][]{randomforest}, Probabilistic Random Forest \citep[PRF][] {probrandfor}, K-Neighbours \citep[KN][]{kneighbours}, Support Vector Regression \citep[SVR][]{svrmethod}, Multi-Layer-Perceptron \citep[MLP][] {mlpmethod}. In the end we identified the SVR method to have the best results, outperforming the other methods by a considerable margin. 

\begin{table*}
\caption{Best photometric input variables selected for each ML stellar parameter estimation (T$_{\rm{eff}}$, log$g$, and [Fe/H]) and for each quality level (gold, silver, bronze). We also indicate the sample size available for both the training and testing of the ML algorithm. See Section~\ref{sec:mlpars} for details.}
\label{tab:mlinput} 
\centering
\begin{tabular}{lllll}
\hline \hline
Dataset & Parameter      & Sample size & Input variables$^*$ \\
\hline
Gold    & T$_{\rm{eff}}$ & 2260        & Dist, E(B--V), $\sigma$E(B--V), U$_{\rm{abs}}$, B$_{\rm{abs}}$, V$_{\rm{abs}}$, R$_{\rm{abs}}$, I$_{\rm{abs}}$, U, B, V, $\sigma$V, R, I, $\sigma$I, T$_{\rm{eff}}^{\rm{photo}}$, [Fe/H]$^{\rm{photo}}$ \\
& log$g$         & 2260        & Dist, U$_{\rm{abs}}$, R$_{\rm{abs}}$, I$_{\rm{abs}}$, U, B, V, R, I, T$_{\rm{eff}}^{\rm{photo}}$, log$g^{\rm{photo}}$, [Fe/H]$^{\rm{photo}}$ \\
& [Fe/H]         & 2366        & Dist, E(B--V), U$_{\rm{abs}}$, B$_{\rm{abs}}$, V$_{\rm{abs}}$, R$_{\rm{abs}}$, B, V, R, I, T$_{\rm{eff}}^{\rm{photo}}$ \\
Silver  & {\em (all)}    & 3154        & Dist, E(B--V), U$_{\rm{abs}}$, B$_{\rm{abs}}$, V$_{\rm{abs}}$, B, V, I, T$_{\rm{eff}}^{\rm{photo}}$ \\
Bronze  & {\em (all)}    & 4290        & Dist, E(B--V), B$_{\rm{abs}}$, V$_{\rm{abs}}$, B, V, I, T$_{\rm{eff}}^{\rm{photo}}$ \\
\hline  
\multicolumn{4}{l}{\small{$^*$The symbols U$_{\rm{abs}}$, B$_{\rm{abs}}$, V$_{\rm{abs}}$, R$_{\rm{abs}}$, I$_{\rm{abs}}$ refer to the absolute and dereddened magnitudes; the "photo" superscript to photometric parameters.}} \\
\end{tabular}
\end{table*}

\begin{table}
\caption{Typical (median) error on the ML estimates of the parameters for each training dataset (gold, silver and bronze).}
\label{tab:mlperf} 
\centering
\begin{tabular}{lrrr}
\hline\hline
Dataset & T$_{\rm{eff}}$ & log$g$ & [Fe/H] \\
          & (K) & (dex) & (dex) \\
\hline
Gold & 98 & 0.14 & 0.14 \\
Silver & 137 & 0.16 & 0.20 \\
Bronze & 403 & 0.59 & 0.45 \\
\hline
\end{tabular}
\end{table}

We initially identified a set of input values for the ML algorithms from the photometric variables present in the catalogue. ML methods cannot correctly treat empty (null or NaN) values, thus the major limiting factor on the size of the sample for training and testing (and on the final number of parametrized stars) is that some parameters, mostly R and U magnitudes, are frequently missing. By minimizing the error, i.e., the difference between the training parameters and the output ones in the test phase, we identified an optimal set of input variables, depending on the specific parameter that we are trying to estimate and that are indicated in Table~\ref{tab:mlinput}. The estimates produced with these input parameters sets are labelled with {\tt starMethod=ml\_gold} in the catalogue (55231 stars). We also identified two reduced sets of input variables, which allow the method to be applied to additional sets of stars, although with reduced accuracy. These samples are labelled as ({\tt starMethod=ml\_silver} and {\tt ml\_bronze}) in the combined catalogue, amounting to 102959 and 1075 stars, respectively. In the case of silver parameters we essentially drop the R magnitude, while for the bronze ones we also drop the U magnitude.

After multiple trials, we selected the following operational strategy to estimate the parameters over the widest possible sample. Since the sample selection can influence the output of the algorithm, we decided to iterate the algorithm 100 times, each time randomly selecting half of the ML sample for training, and the other half for testing. Then, for each iteration, we applied the method to the whole sample of stars having all the relevant input variables. We selected as our best estimate for each star the median value of the 100 parameter estimates. We verified that the result is consistent with the best possible estimate using a single iteration and 100\% of the ML sample for training. This approach permits to compute a Median Absolute Deviation (MAD) of the estimate for each star, which can be seen as the precision of the trained method in producing the same value independently of the sample selection. This also means that the method is self-consistent and reproducible within the MAD, and we can use it to filter out predictions with larger MAD values as inaccurate. For each parameter, we also computed the standard deviation of the previously defined best estimate on the ML sample by comparing with the spectroscopic parameters (Table~\ref{tab:mlperf}, see also Figure~\ref{fig:parcomp}). We finally computed the uncertainty on each parameter and for each star by taking the quadrature sum of the MAD (on each star) and the above standard deviation.

In Figure~\ref{fig:parcomp} (bottom panels) we show the error distribution obtained for each estimated variable obtained over the ML sample (as described before). We highlight that the [Fe/H] estimates obtained with SVR, differently from what observed in \citet{anders21} and \citet{miller15}, do not show such a large trend of overestimation for low metallicity and underestimation for high metallicity. Also, there is no saturation effect in the ML metallicity, as opposed to the estimates by \citet{chiti21}. We believe that the main reason for the better performance obtained here lies mostly in the choice of the algorithm. In fact, when using a Random Forest insted of the SVR, we found the same trend as \citet{anders21} and \citet{miller15} with the spectroscopic metallicity. Other factors that may have also helped are the choice of a large and diverse set of input parameters (see Table~\ref{tab:mlinput}) and the fact that we used a reliable and large sample of spectroscopic parameter estimates for the training of the algorithm\footnote{The spectroscopic data employed here come for the most part from the SoS \citet{tsantaki21}. It was shown by \citet{soubiran21} that the surveys tend to slightly overestimate low metallicities and underestimate the high ones compared to higher resolution studies. That effect goes in the same direction of the slope visible in the top-right panel of Figure~\ref{fig:parcomp}, but is much smaller, of the order of 0.1--0.2~dex.}, covering the relevant parameter space. We believe it will be worth investigating the matter further and trying to apply our method to larger photometric samples such as the SDSS or Pan-STARRS in the near future.

Finally, we filtered out the following samples:

\begin{itemize}
    \item{white dwarfs estimates are particularly unrealistic in all parameters ($\simeq$ 800 stars);}
    \item{some of the [Fe/H] results outside of the range covered by the training sample (--3 to +1~dex), and with a MAD greater than 0.9~dex ($\simeq$3000 stars);}
    \item{most of the log$g$ results below --0.5~dex, which had MAD higher than 0.9~dex ($\simeq$1600 stars);}
    \item{T$_{\rm{eff}}$ for stars with (V--I)$_0<-0.23$~mag, which appear to be too cool by hundreds or thousands of degrees, perhaps due to an under-representation of the category in the ML sample ($\simeq$300 stars).}
\end{itemize}




\begin{figure}[t]
    \centering
    \resizebox{\hsize}{!}{\includegraphics[clip]{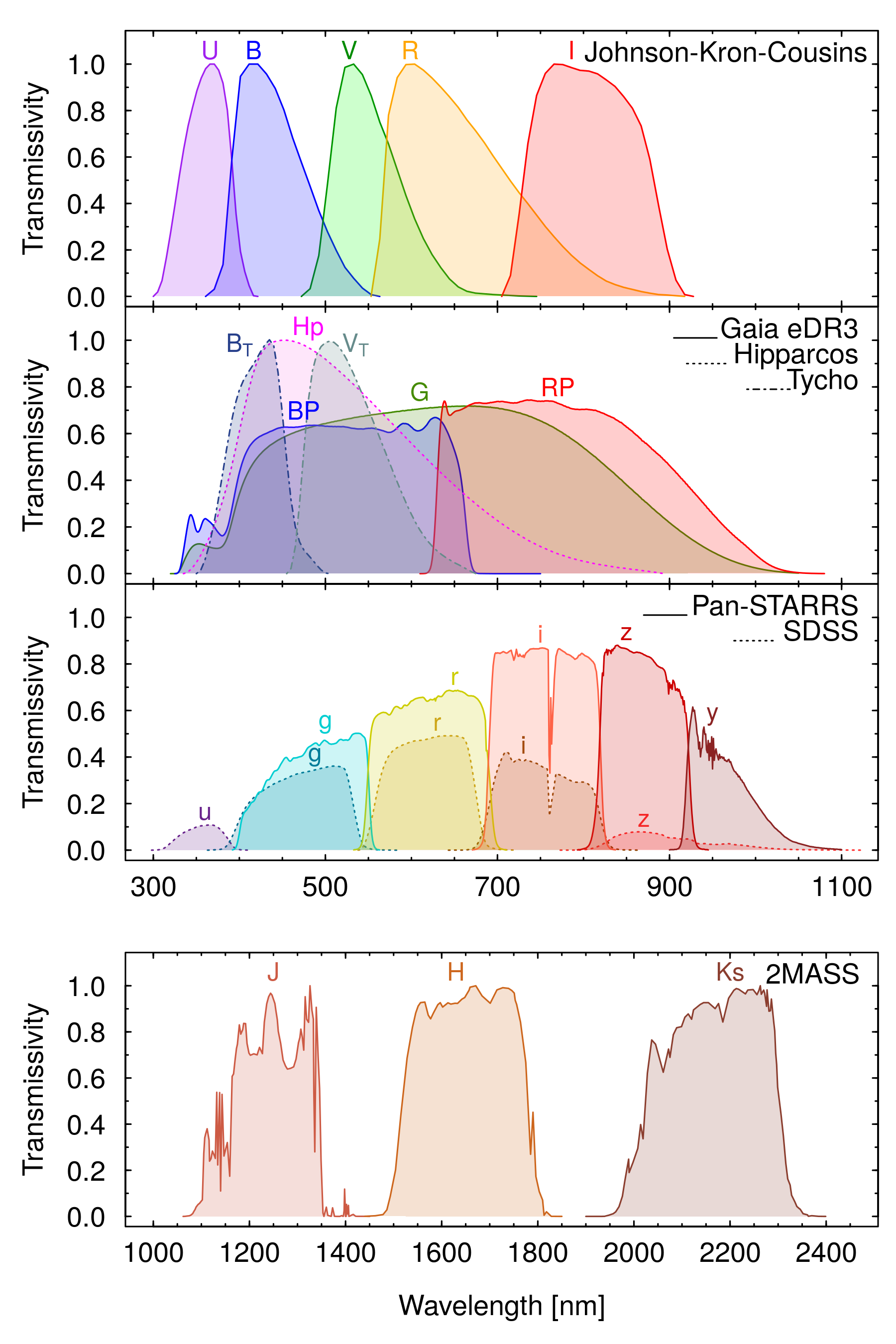}}
    \caption{Summary of the photometric systems used to compute color transformations with the Johnson-Kron-Cousins system. The passbands were obtained from  the SVO Filter Profile Service (see Section~\ref{sec:trans} for more details).}      \label{fig:bands}
\end{figure}

\section{Relations with other photometric systems}
\label{sec:trans}

To obtain transformations between the Johnson-Kron-Cousins system and some of the currently most used photometric systems, we looked for the counterparts of our clean sample standards (Section~\ref{sec:gold}) in the large photometric surveys that use those photometric systems. To this aim, we used the official cross-matches of {\em Gaia} ERD3 with large photometric surveys, publicly available at the SSDC {\em Gaia} portal\footnote{\url{http://gaiaportal.ssdc.asi.it/}} or the {\em Gaia} archive\footnote{\url{https://gea.esac.esa.int/archive/}}, which were obtained with the same cross-match algorithm used here \citep{marrese17,marrese19}. Moreover, we only selected stars with a relatively low reddening, E(B--V)$\lesssim$0.3--0.8~mag, depending on the exact color combination and the direction of the reddening vector in that plane. The final catalogue of the magnitudes of the Landolt and Stetson standards in all the considered photometric systems is presented in Table~\ref{tab:mags}. More details about the sample selections can be found in the following sections. Figure~\ref{fig:bands} summarizes the photometric systems and passbands considered. The transmissivity curves used in the figure were obtained from the SVO Filter Profile Service\footnote{\url{http://svo2.cab.inta-csic.es/theory/fps/}} \citep{svo1,svo2}. In particular, we used the passbands by \citet{bessell12}, \citet{riello20}, \citet{moro00}, \citet{tonry12}, \citet{doi10}, and \citet{cohen03}.

For each photometric system, we computed the transformations as polynomial fits in the form: \begin{equation} Y = \Sigma_i~ a_i ~ X^i \label{eq:edr3} \end{equation} where $Y$ is a difference between two magnitudes in the two systems  (e.g., G--V or B--G) and $X$ is a color in one of the two systems (e.g., B--V or G$_{\rm{BP}}$--G$_{\rm{BP}}$). The choice of the order of the polynomial was made by inspecting the residual plots and by selecting the lowest polynomial order that removes any residual systematic oscillations (wavy patterns) in the residuals, or that reduces it below the overall spread in the residuals. Although the data are often of very high quality, there are sometimes substructures, jumps, or secondary branches. These arise naturally because of: {\em (i)} the magnitude dependent uncertainties and {\em (ii)} the different sensitivity of the employed photometric passbands to different spectral features. Rather than using a weighted fit, which would bias the fits in favour of bright stars, we opted for a smoothing of the Y vector before fitting. We replaced each Y value with the median of the three closest values after sorting in $X$, iteratively for three times \citep{tukey77}. This minimizes any outlying features caused by higher reddening stars or specific minority stellar sub-groups (such as WDs or SDs), and it also has the side advantage of slightly improving the fits at the borders, where there are statistically much fewer stars. The disadvantage is that minority components (e.g., WDs or SDs) can be in some cases not well represented by the fits. The coefficients of all the 167 polynomial fits are presented in Table~\ref{tab:trans}. More details on the fitting procedure, adopted choices, and goodness of the fits can be found in the following sections.

With the data presented in Table~\ref{tab:mags}, it is possible to compute additional color transformations between the explored samples in color combinations that we did not include in Table~\ref{tab:trans}. Using the data in Table~\ref{tab:cata}, it should also be relatively easy to identify Landolt and Stetson standards in any other survey or catalogue, if they are present. This will enable the computation of transformations to and from additional photometric systems not considered in this work. Finally, we have used mostly giants and dwarfs, so in several cases our transformations do not apply to WDs, YSO, or hot subdwarfs. However, with the classification provided in Table~\ref{tab:cata} and the magnitudes in Table~\ref{tab:mags}, users can derive their own specific transformations for these and other less represented stellar types.

\begin{table}
\caption{Magnitudes of stars in the combined catalogue (Table~\ref{tab:cata}) in the different photometric systems considered in Section~\ref{sec:trans}. Only stars with a {\em Gaia} EDR3 match, belonging to the clean sample (Section~\ref{sec:gold}), and with a match in at least one of the photometric surveys, after survey-specific quality cuts, are included (see Sections~\ref{sec:edr3}--\ref{sec:2mass} for details).}
\label{tab:mags} 
\centering                         
\begin{tabular}{lll}        
\hline\hline                
Column & Units & Description \\  
\hline                       
ID & & Star ID from Table~\ref{tab:cata} \\  
Gaia EDR3 ID & & Gaia Source ID from EDR3 \\ 
G$^\prime$ & (mag) & Corrected {\em Gaia} G magnitude \\
$\delta$G$^\prime$ & (mag) & Recomputed error on G$^\prime$ \\  G$_{\rm{BP}}^\prime$ & (mag) & Corrected {\em Gaia} G$_{\rm{BP}}$ magnitude \\
$\delta$G$_{\rm{BP}}^\prime$ & (mag) & Recomputed error on G$_{\rm{BP}}^\prime$ \\
G$_{\rm{RP}}^\prime$ & (mag) & Corrected {\em Gaia} G$_{\rm{RP}}$ \\
$\delta$G$_{\rm{RP}}^\prime$ & (mag) & Recomputed error on G$_{\rm{RP}}^\prime$ \\
HIP ID & & HIPPARCOS identifier \\
H$_{\rm P}$           & (mag) & Hipparcos magnitude \\
$\delta$H$_{\rm P}$   & (mag) & Error on H$_{\rm P}$ \\
TYC ID & & Tycho identifier \\
B$_{\rm T}$     & (mag) & Tycho B magnitude \\
$\delta$B$_{\rm T}$ & (mag) & Error on B$_{\rm T}$ \\
V$_{\rm T}$     & (mag) & Tycho V magnitude \\
$\delta$V$_{\rm T}$ & (mag) & Error on V$_{\rm T}$ \\
SDSS ID & & SDSS DR13 identifier \\
u$_{\rm{SDSS}}$ & (mag) & u PSF magnitude in SDSS\, DR13 \\
$\delta$u$_{\rm{SDSS}}$ & (mag) & Error on u$_{\rm{SDSS}}$ \\
g$_{\rm{SDSS}}$ & (mag) & g PSF magnitude in SDSS\, DR13 \\
$\delta$g$_{\rm{SDSS}}$ & (mag) & Error on g$_{\rm{SDSS}}$ \\
r$_{\rm{SDSS}}$ & (mag) & r PSF magnitude in SDSS\, DR13 \\
$\delta$r$_{\rm{SDSS}}$ & (mag) & Error on r$_{\rm{SDSS}}$ \\
i$_{\rm{SDSS}}$ & (mag) & i PSF magnitude in SDSS\, DR13 \\
$\delta$i$_{\rm{SDSS}}$ & (mag) & Error on i$_{\rm{SDSS}}$ \\
z$_{\rm{SDSS}}$ & (mag) & z PSF magnitude in SDSS\, DR13 \\
$\delta$z$_{\rm{SDSS}}$ & (mag) & Error on z$_{\rm{SDSS}}$ \\
PS1 ID & & Pan-STARRS-1 identifier \\
g$_{\rm{PS}}$ & (mag) & g PSF magnitude in PS1 \\
$\delta$g$_{\rm{PS}}$ & (mag) & Error on g$_{\rm{PS}}$ \\
r$_{\rm{PS}}$ & (mag) & r PSF magnitude in PS1 \\
$\delta$r$_{\rm{PS}}$ & (mag) & Error on r$_{\rm{PS}}$ \\
i$_{\rm{PS}}$ & (mag) & i PSF magnitude in PS1 \\
$\delta$i$_{\rm{PS}}$ & (mag) & Error on i$_{\rm{PS}}$ \\
z$_{\rm{PS}}$ & (mag) & z PSF magnitude in PS1 \\
$\delta$z$_{\rm{PS}}$ & (mag) & Error on z$_{\rm{PS}}$ \\
y$_{\rm{PS}}$ & (mag) & y PSF magnitude in PS1 \\
$\delta$y$_{\rm{PS}}$ & (mag) & Error on y$_{\rm{PS}}$ \\
2MASS ID & & 2MASS PSC identifier \\
J$_{\rm{2MASS}}$ & (mag) & J magnitude from 2MASS PSC \\
$\delta$J$_{\rm{2MASS}}$  & (mag) & Error on J$_{\rm{2MASS}}$ \\
H$_{\rm{2MASS}}$ & (mag) & H magnitude from 2MASS PSC \\
$\delta$H$_{\rm{2MASS}}$  & (mag) & Error on H$_{\rm{2MASS}}$ \\
K$_{\rm{2MASS}}$ & (mag) & K$_s$ magnitude from 2MASS PSC \\
$\delta$K$_{\rm{2MASS}}$  & (mag) & Error on K$_{\rm{2MASS}}$ \\
\hline                      
\end{tabular}
\end{table}

\begin{table}
\caption{Coefficients of the polynomial fits (Equation~\ref{eq:edr3}) to convert magnitudes between the Johnson-Kron-Cousins photometric system and literature photometric systems of various photometric surveys, based on the data in Table~\ref{tab:mags}. See Section~\ref{sec:trans} for more details.}
\label{tab:trans} 
\centering                         
\begin{tabular}{lll}        
\hline\hline                
Column & Units & Description \\  
\hline                       
Survey    &       & Provenance of the survey data \\
          &       & (example: Gaia EDR3, SDSS DR13, ...) \\
System    &       & Photometric system of the survey data \\
          &       & (example: Gaia, ugriz, JHK, ...) \\
X         & (mag) & Input color (examples: V--I or G$_{\rm{BP}}$--G$_{\rm{RP}}$) \\
X$_{\rm{min}}$ & (mag) & Lower validity limit for the fit \\
X$_{\rm{max}}$ & (mag) & Upper validity limit for the fit \\
Y         & (mag) & Output color (examples: G--V or g--V) \\
$a_0$     & & Coefficient of zero-order term \\
$a_1$     & & Coefficient of first-order term (X) \\
$a_2$     & & Coefficient of second-order term (X$^2$) \\
...       & & (i.e., one column for each coefficient) \\
$a_n$     & & Coefficient of nth-order term (X$^n$) \\
$\sigma$  & (mag) & Standard deviation of Y residuals \\
N         & & Number of stars used in the fit \\
Validity     & & Notes on validity regime \\
          & & (examples: giants, dwarfs, all) \\
Notes & & Additional cautionary notes \\
          & & (example: not for WDs) \\
\hline                      
\end{tabular}
\end{table}

\begin{figure}[t]
    \centering
    \resizebox{\hsize}{!}{\includegraphics[clip]{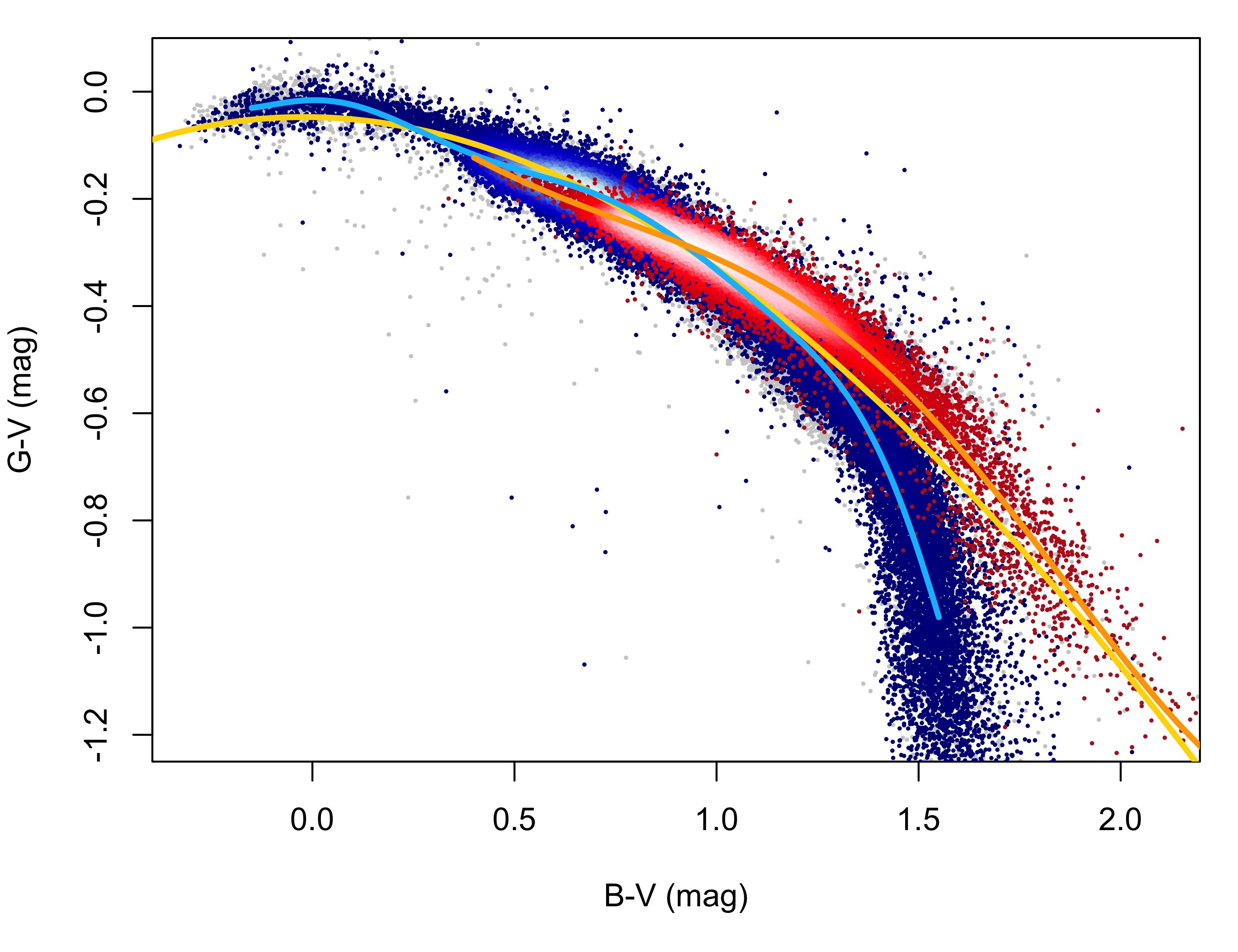}}
    \caption{Example of transformation from V, B--V to the {\em Gaia} G magnitude, illustrating the advantage of performing separate polynomial fits for dwarfs and giants. The whole clean sample is plotted in the background as grey dots, dwarfs are plotted in blue, and giants in red. The polynomials fitted on the two samples separately are plotted as a light blue and an orange line, respectively. The polynomial fit by \citet{riello20} is plotted in yellow: it reproduces the dwarfs at blue colors, the giants at red colors, and passes between the two at intermediate colors.}
    \label{fig:split}
\end{figure}

\subsection{The Gaia EDR3 system}
\label{sec:edr3}

Before computing the transformations between the Johnson-Kron-Cousins and the {\em Gaia} EDR3 photometric system (Figure~\ref{fig:bands}), we corrected for some known effects in the {\em Gaia} EDR3 photometry. This has to be taken into consideration when using our transformations on the publicly available {\em Gaia} EDR3 magnitudes. In particular:

\begin{itemize}
    \item{saturated stars were corrected following the equations C.1, C.2, and C.3 in Appendix~C by \citet{riello20}; although the impact of saturation has decreased with respect to {\em Gaia} DR2 \citep{rowell21}, a small correction is still needed;}
    \item{the G magnitude of stars with a six-parameter solution, i.e., for which  the BP and RP magnitudes where not available to produce the astrometry and, most importantly the PSF/LSF calibration \citep{rowell21}, were corrected following the recipe in Section~8.3 and Table~5 by \citet{riello20};}
    \item{the G magnitudes of bright blue giants (8$<$G$<$13\,mag and G$_{\rm{BP}}$--G$_{\rm{RP}}$$<$--0.1\,mag) has been corrected following the recipe in Section~8.4 by \citet{riello20}; the linear trend with G is probably caused by small problems in the PSF/LSF calibration \citep{rowell21}.}
\end{itemize}

The $a_i$ coefficients for 35 different {\em Gaia} and Johnson-Kron-Cousins color combinations are provided in Table~\ref{tab:trans}, along with their color validity ranges, the standard deviation of the $Y$ residuals of each fit, and relevant annotations. An example is presented in Figure~\ref{fig:split}, where a comparison with the transformations by \citet{riello20} is also presented. Outside of the indicated color validity ranges, the transformations are either less accurate or completely unreliable, depending on the polynomial behavior at the borders, and therefore caution is required. In particular, we noted that in most of the relations reported in Table~\ref{tab:trans}, it was necessary to perform separate fits for dwarf and giant stars. The largest differences are observed for M stars, as already noted by \citet{riello20}, but in most cases dwarfs and giants occupy different loci at all colors, not just at the reddest ones, as illustrated in Figure~\ref{fig:split}. We separated giants and dwarfs using both our stellar classification (Section~\ref{sec:cmd}) and parametrization (Section~\ref{sec:sos}). In particular, we separated the samples at log$g$=3.5~dex and we further removed stars with log$g$$>$5.5~dex from the dwarfs sample (except for those cases in which the WDs and SDs could be well fitted together with the dwarfs). We also note that, to improve the fits, we removed highly reddened stars, therefore our transformations are in general not appropriate for stars with reddening E(B--V)$\gtrsim$0.5~mag. 

The WDs and hot sub-dwarfs did not follow the same relations as the normal dwarfs and giants in some of the fits. When this occurred, we noted it down in the {\tt Notes} column of Table~\ref{tab:trans}. Additionally, some transformations involving the U band, which is not well covered by the {\em Gaia} passbands, require a very high order and provide large spreads (3\% or more); some others have a non monotonic behaviour\footnote{This is expected when involving the U band, or more in general when using bands lying on the opposite sides of the Balmer jump, but it was also observed in bands not involving the Balmer jump.}. In some of these cases, we still provided the coefficients, but inserted a warning in the {\tt Notes} column. Finally, while the standard U band can be generally used to predict {\em Gaia} photometry, the opposite is not true, because a simple polynomial cannot be used to reproduce the complex behaviour in these planes with the G$_{\rm{BP}}$--G$_{\rm{RP}}$ color and magnitude differences involving the U band. 

\subsection{The Hipparcos and Tycho systems}
\label{sec:hip}

The Hipparcos astrometric satellite \citep{perryman97,hog97,vanleeuwen07}\footnote{\url{https://www.cosmos.esa.int/web/hipparcos/catalogues}}, the predecessor of {\em Gaia}, provided photometry in one very wide, white-light band labelled H$_{\rm P}$. The photometric system has been discussed in detail by \citet{bessell00}. The Tycho instrument onboard also provided color information, by means of a blue and a red wide band, B$_{\rm T}$ and V$_{\rm T}$, respectively (Figure~\ref{fig:bands}). We used the cross-match between {\em Gaia} EDR3 \citep{marrese17,marrese19} with the re-reduction of the original Hipparcos data \citep{vanleeuwen07} and the Tycho-2 re-analysis of the Tycho data \citep{hog00}. We found 76 stars in common with the Hipparcos and 1315 with the Tycho catalogue.

Given the small sample sizes and the large uncertainties in the B$_{\rm T}$ and V$_{\rm T}$ magnitudes, the spreads on the computed relations, especially those involving the B$_{\rm T}$ and V$_{\rm T}$ bands, are often of 10--20\%, reaching above 30\% when involving the U band. The two transformations involving only H$_{\rm P}$ have instead spreads of the order of 2--3\%. We did not perform separate fits for dwarfs and giants. We computed 10 relations between H$_{\rm P}$ and the $UBVRI$ magnitudes, using the smaller sample of Hipparcos stars with simultaneous estimates of H$_{\rm P}$, B$_{\rm T}$, and V$_{\rm T}$ (these transformations are labeled "Hipparcos 2007" in Table~\ref{tab:trans}). We also computed 10 relations between the $UBVRI$ and the B$_{\rm T}$, and V$_{\rm T}$ bands, using the larger sample of Tycho-2 stars (these are labelled "Tycho-2" in Table~\ref{tab:trans}), where we selected only stars with $\delta$B$_{\rm{T}}$ and $\delta$V$_{\rm{T}}$ smaller than 0.1~mag for the fits, remaining with 24 stars from Hipparcos and 786 for Tycho. The polynomial orders of these fits are considerably lower than in other systems, between 2 and 3.

\begin{figure}[t]
    \centering
    \resizebox{\hsize}{!}{\includegraphics[clip]{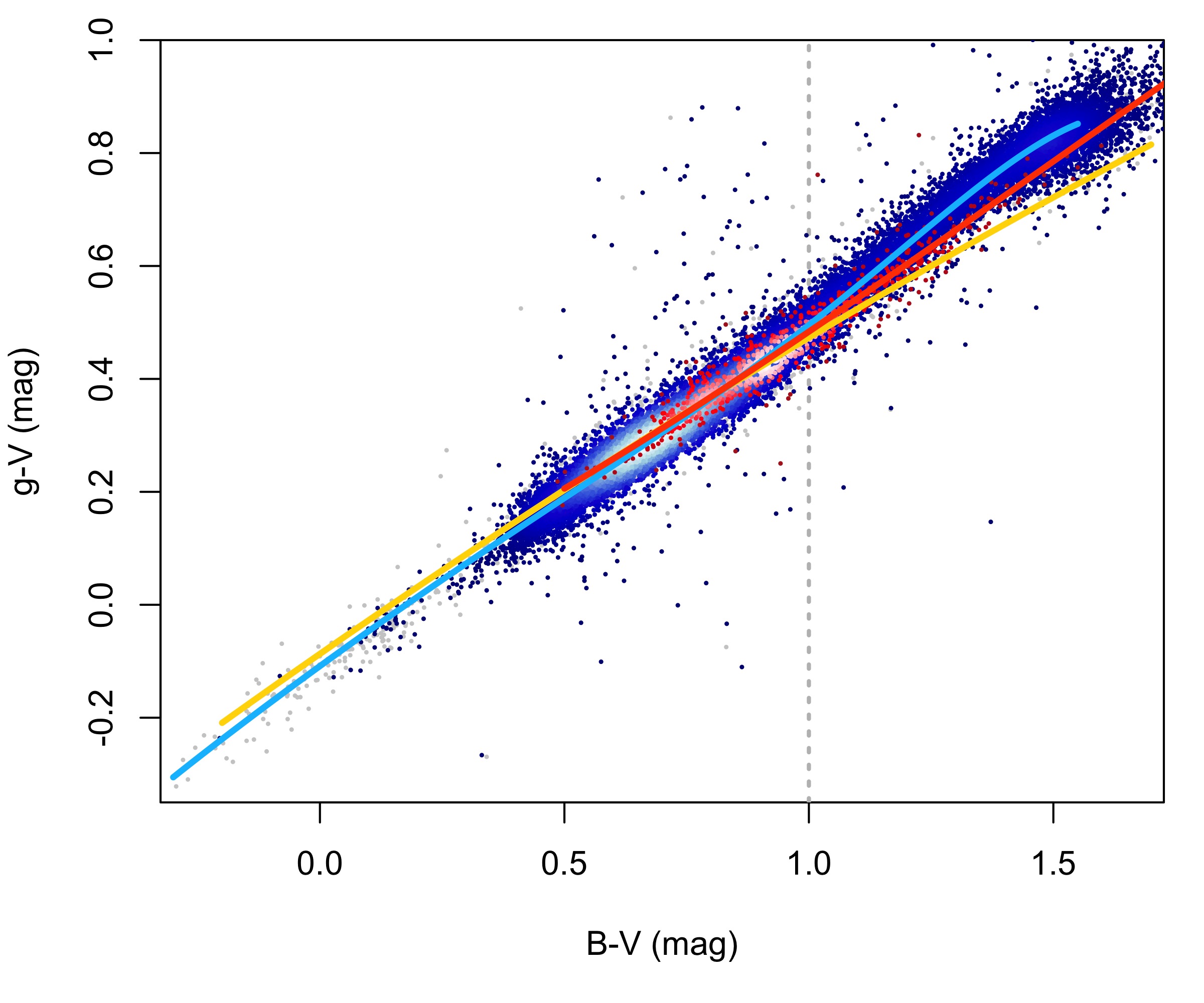}}
    \caption{Similar to Figure~\ref{fig:split}, but for the SDSS DR13 ugriz system. In particular, the B--V, g--V case is shown. The clean sample is plotted in the background as grey dots, dwarfs are plotted in blue, and giants in red. The polynomials fitted on the two samples separately are plotted as a light blue and an orange line, respectively. The original transformation by \citet{rodgers06}, who used the \citet{smith02} ugriz standards and who could only count on four stars redder than B--V=1~mag (grey dotted line), is shown in yellow. For B--V$<$1~mag, the three relations agree within about 2\%. In general, the \citet{rodgers06} relation agrees more with the giants sample than with the dwarfs one.}
    \label{fig:sdss}
\end{figure}

\subsection{The SDSS DR13 ugriz system}
\label{sec:sdss}

The ugriz photometric system was first defined by \citet{fukugita96} and \citet{newberg99} and it is the system of choice to derive photometric redshifts of (faint) external galaxies for cosmological studies, thanks to its very wide and equally spaced passbands (Figure~\ref{fig:bands}), thus it does not focus on specific stellar features \citep{bessell05}. Initially, the passbands were indicated in the literature with apices (u$^\prime$g$^\prime$r$^\prime$i$^\prime$z$^\prime$), to discern them from those of the older \citet{thuan76} system, which had narrower bands and no z filter. We will not use apices in the following. The standard stars in the ugriz system were provided by \citet{smith02} and the tranformations with the Johnson-Kron-Cousins system were provided by \citet{rodgers06}. It is worth mentioning here that the Vera Rubin telescope\footnote{\url{https://www.lsst.org/}} \citep[previously known as LSST,][]{ivezic19} will instead employ a photometric system similar to the SDSS ugriz and Pan-STARRS grizy ones, but slightly different from both in the passband shapes. Additionally, APASS \citep{apass1,apass2}, which is widely used by the community studying variable stars, uses the SDSS griz filters in combination with the Johnson-Kron-Cousins BV bands.

To derive new transformations using our standard catalogue, we used the publicly available {\em Gaia} EDR3 cross-match \citep{marrese17,marrese19} with SDSS DR13 \citep{albareti17}. In that release, a complete re-assessment of the SDSS half-billion object photometry was performed, including improvements of the data processing and a full re-calibration of the ugriz magnitudes \citep{finkbeiner16}, based on the zero-points of the Pan-STARRS system by \citet{schlafly12}\footnote{The original zero-point of the SDSS ugriz system was set using BD\,+17\,4708, which is also one of the {\em Gaia} candidate spectro-photometric standard stars \citep[i.e., SPSS\,20,][]{pancino12}. This star was shown to be variable, with an amplitude of 65~mmag, by \citet{marinoni16} and was thus rejected as a standard for {\em Gaia}.}. We used the SDSS {\tt clean} flag to select high-quality measurements. We excluded all duplicates found by the cross-match and we removed all stars with more than one neighbor. We further removed stars with E(B--V)$\gtrsim$0.3--0.8~mag, depending on the case, and errors on the SDSS magnitudes above the 95\% percentile, as a function of magnitude, in at least three bands. We remained with a sample of about $\simeq$38\,000 stars out of the original $\simeq$63\,000. 

We used Equation~\ref{eq:edr3} and computed 26 polynomial fits to obtain SDSS magnitudes from Johnson-Kron-Cousins ones and vice versa. We restricted the relations to bands covering similar wavelength ranges (Figure~\ref{fig:bands}) and therefore we could often use polynomials of lower order, especially for giants. Similarly to the previous cases, we found that it is often more effective to perform separate fits for dwarfs and giants. This occurrence is noted down in the {\tt Validity} column of Table~\ref{tab:trans}. Additionally, transformations involving the blue bands are generally not appropriate for cool stars and those obtained from red bands are less accurate for the hot stars. The transformations are also not appropriate for highly reddened stars, or for WD and hot sub-dwarfs in general. 

An example of polynomial fit for the SDSS ugriz system is shown in Figure~\ref{fig:sdss},  where we also compare with the original transformations from \citet{rodgers06}. It is important to note that the \citet{rodgers06} transformations were based on the \citet{smith02} standards and that only four stars were present in that sample with colors redder than B-V=1~mag. For this reason, their relation and ours diverge for redder stars in Figure~\ref{fig:sdss}. However, the agreement for B--V$<$1~mag is $\leq$2\%. 

\subsection{The Pan-STARSS-1 grizy system}
\label{sec:panstarss}

The Panoramic Survey Telescope and Rapid Response System \citep[Pan-STARRS,][]{ps1_1,ps1_2} photometric system \citep{tonry12,magnier20} is similar to the ugriz one adopted by SDSS, but it does not have u and it adds a redder y band. At the same time, the passband shapes are slightly different from the SDSS ones (see Figure~\ref{fig:bands}) and small differences are also expected with the upcoming Vera Rubin telescope passbands \citep{ivezic19}. 

Similarly to the other surveys, we used the official {\em Gaia} cross-match \citep{marrese17,marrese19} to look for stars in common with our sample. For the reference photometry, we used the pre-filtered version of the PS1 release available in the {\em Gaia} archive, where data with only one detection, with large or null uncertainties or magnitudes, and with large errors on the coordinates were removed. We further removed stars with more than one neighbor or mate. This provided us with a starting clean sample of about 88\,000 stars in common with our combined catalogue. Similarly to the case of the SDSS in the previous section, we computed 22 transformations, using the closest passbands in wavelength, therefore we did not need to use often high-degree polynomials. We also separated the giants from the dwarfs in most cases, and the WDs generally were well reproduced by the transformations appropriated for the dwarfs, unlike in the case of SDSS.

\subsection{The 2MASS JHK system}
\label{sec:2mass}

The 2MASS survey \citep[Two-Micron All Sky Survey][]{2mass1,2mass2} is among the most used photometric  survey in the near infrared wavelength range. Several authors provide transformations between optical photometric systems and the JHK system used in 2MASS. We believe that, because of the two very different wavelength regimes covered, these transformations have to be used with great caution, but we do understand that they can be helpful in a variety of applications. Here, we use the 2MASS PSC (Point Source Catalogue) of 470 million sources, which has been cross-matched with {\em Gaia} EDR3 using the same software used elsewhere in this paper \citep{marrese17,marrese19}. We selected stars with no mates and no neighbours in the cross-match table, and we further imposed an angular distance smaller than 0.5". We also made use of the various flags in the 2MASS PSC, to remove sources affected by blends, projected on extended sources, or with problems in the photometric data or processing. From a list of $\simeq$150\,000 stars in common with our combined catalogue, we thus selected the best $\simeq$90\,000. 

We did not provide any transformations with the U-band, because of the large scatter around the fits ($>$0.5~mag) and the large dependency on metallicity. We found, in the U-band fits, color differences between solar metallicity and metal-poor stars reaching up to one magnitude, especially for giants. The dependency on metallicity is present also in other bands, mostly when involving  the B, R, or H bands. Unfortunately, for the dwarfs sample, we could not count on enough good metallicity estimates, thus our fits for dwarfs are only valid for stars with [Fe/H]$\gtrsim$--1.0~dex. For giants, we sometimes split the sample at [Fe/H]=--0.8~dex and performed separate fits for the two groups, as appropriate and as annotated in the {\tt Validity} column of Table~\ref{tab:trans}. We computed in total 63 transformations with the 2MASS JHK system. In general, the transformations to obtain Johnson-Kron-Cousins magnitudes from 2MASS ones have large spreads, around 0.3--0.4, 0.2--0.3, 0.1--0.2 and 0.1~mag for B, V, R, and I respectively. The transformations to get 2MASS magnitudes from Johnson-Kron-Cousins ones have generally lower spreads, ranging from 0.03~mags to 0.1~mags.

An additional complication is that the different wavelength range covered (optical vs. infrared) implies a large amount of substructures in the color plots, revealing important differences in the underlying spectral energy distribution of the stars. For example, several bifurcations are evident in certain color combinations, which split stars according to their reddening (even if we remained below E(B--V)$\simeq$0.5~mag), to their gravity, separating giants from subgiants, and to their metallicity. Following all of these bifurcations would have multiplied the number of transformations that we had to provide, so we generally discarded the minority branches and provided the relevant information in the notes in Table~\ref{tab:trans}. However, the potential -- in terms of stellar classification power -- of a combined optical-infrared photometry is enormous and is definitely worth exploring further.


\section{Conclusions}
\label{sec:concl}

We have provided a curated version of the widely used collections of secondary standards provided by Landolt and Stetson and their collaborators. We have selected stars with comparatively higher photometric accuracy and we profited from {\em Gaia} EDR3 data and literature catalogues to flag confirmed or suspected blends, binaries, and variable stars. We also classified and characterized stars making a wide use of the available literature data and of the exquisite {\em Gaia} data, in terms of evolutionary stage, stellar types, atmospheric parameters, reddening, distance, and radial velocity. In particular, we used the SVR regression algorithm, trained on the SoS spectroscopic data \citep{tsantaki21} to derive machine learning parameters from the Landolt and Stetson photometry, which provides significantly better results compared to similar machine learning studies in the literature, especially as far as log$g$ and [Fe/H] determinations are concerned. We used the {\em Gaia} cross-match algorithm to identify the high-quality sample of Landolt and Stetson secondary standards in external surveys ({\em Gaia}, Hipparchos and Tycho, SDSS, Pan-STARRS-1, and 2MASS) and computed 167 transformations between those photometric systems and the Johnson-Kron-Cousins one. All the data presented in the paper tables are available electronically, on the CDS Vizier catalogue service and on the Gaia Portal at SSDC (\url{http://gaiaportal.ssdc.asi.it/}).


\begin{acknowledgements} {\bf People.} We acknowledge the following colleagues for discussions, help, and ideas: F.~De Angeli, M.~Bellazzini, R.~Contreras, D.~W.~Evans, M.~Fabrizio, G.~Fanari, M.~Manteiga, C.~E.~Mart\'\i nez-V\'{a}zquez, P.~Montegriffo, R.~M.~Murillo, M.~Riello, M.~Salvato, P.~B.~Stetson, M.~Zoccali. We would also like to thank M.~Bessell, who was the referee of this and other papers from our team on the subject, for helping us improving the paper and for his constructive and helpful attitude.
{\bf Funding.} This research has been partly funded through the INAF Main Stream grant 1.05.01.86.22 {\em "Chemo-dynamics of globular clusters: the Gaia revolution"} (PI: Pancino). We also acknowledge the financial support of ASI (Agenzia Spaziale Italiana) under the contract to INAF: ASI 2014-049-R.0 dedicated to SSDC, and under the contracts to INAF: ARS/96/77,ARS/98/92, ARS/99/81, I/R/32/00, I/R/117/01, COFIS-OF06-01,ASI I/016/07/0, ASI I/037/08/0, ASI I/058/10/0, ASI 2014-025-R.0, ASI 2014-025-R.1.2015, ASI 2018-24-HH.0, dedicated to the Italian participation to the Gaia Data Analysis and Processing Consortium (DPAC). 

{\bf Data.} This work uses data from the European Space Agency (ESA) space mission {\em Gaia}. {\em Gaia} data are being processed by the Gaia Data Processing and Analysis Consortium (DPAC). Funding for the DPAC is provided by  national institutions, in particular the institutions participating in the {\em Gaia} Multi-Lateral Agreement(MLA). Funding for the Sloan Digital Sky Survey IV has been provided by the Alfred P. Sloan Foundation, the U.S. Department of Energy Office of Science, and the Participating Institutions. SDSS acknowledges support and resources from the Center for High-Performance Computing at the University of Utah. The SDSS web site is www.sdss.org. SDSS is managed by the Astrophysical Research Consortium for the Participating Institutions of the SDSS Collaboration including the Brazilian Participation Group, the Carnegie Institution for Science, Carnegie Mellon University, Center for Astrophysics -- Harvard \& Smithsonian (CfA), the Chilean Participation Group, the French Participation Group, Instituto de Astrofísica de Canarias, The Johns Hopkins University, Kavli Institute for the Physics and Mathematics of the Universe (IPMU) / University of Tokyo, the Korean Participation Group, Lawrence Berkeley National Laboratory, Leibniz Institut für Astrophysik Potsdam (AIP), Max-Planck-Institut für Astronomie (MPIA Heidelberg), Max-Planck-Institut für Astrophysik (MPA Garching), Max-Planck-Institut für Extraterrestrische Physik (MPE), National Astronomical Observatories of China, New Mexico State University, New York University, University of Notre Dame, Observatório Nacional / MCTI, The Ohio State University, Pennsylvania State University, Shanghai Astronomical Observatory, United Kingdom Participation Group, Universidad Nacional Aut\'o noma de M\'e xico, University of Arizona, University of Colorado Boulder, University of Oxford, University of Portsmouth, University of Utah, University of Virginia, University of Washington, University of Wisconsin, Vanderbilt University, and Yale University. The GALAH Survey is based on data acquired through the Australian Astronomical Observatory, under programs: A/2013B/13 (The GALAH pilot survey); A/2014A/25, A/2015A/19, A2017A/18 (The GALAH survey phase 1); A2018A/18 (Open clusters with HERMES); A2019A/1 (Hierarchical star formation in Ori OB1); A2019A/15 (The GALAH survey phase 2); A/2015B/19, A/2016A/22, A/2016B/10, A/2017B/16, A/2018B/15 (The HERMES-TESS program); and A/2015A/3, A/2015B/1, A/2015B/19, A/2016A/22, A/2016B/12, A/2017A/14 (The HERMES K2-follow-up program). We acknowledge the traditional owners of the land on which the AAT stands, the Gamilaraay people, and pay our respects to elders past and present. This paper includes data that has been provided by AAO Data Central (\url{datacentral.org.au}). Guoshoujing Telescope (the Large Sky Area Multi-Object Fiber Spectroscopic Telescope LAMOST) is a National Major Scientific Project built by the Chinese Academy of Sciences. Funding for the project has been provided by the National Development and Reform Commission. LAMOST is operated and managed by the National Astronomical Observatories, Chinese Academy of Sciences. Funding for Rave has been provided by: the Leibniz Institute for Astrophysics Potsdam (AIP); the Australian Astronomical Observatory; the Australian National University; the Australian Research Council; the French National Research Agency; the German Research Foundation (SPP 1177 and SFB 881); the European Research Council (ERC-StG 240271 Galactica); the Istituto Nazionale di Astrofisica at Padova; The Johns Hopkins University; the National Science Foundation of the USA (AST-0908326); the W. M. Keck foundation; the Macquarie University; the Netherlands Research School for Astronomy; the Natural Sciences and Engineering Research Council of Canada; the Slovenian Research Agency; the Swiss National Science Foundation; the Science \& Technology FacilitiesCouncil of the UK; Opticon; Strasbourg Observatory; and the Universities of Basel, Groningen, Heidelberg and Sydney. Based on observations obtained with the Samuel Oschin Telescope 48-inch and the 60-inch Telescope at the PalomarObservatory as part of the Zwicky Transient Facility project. ZTF is supported by the National Science Foundation under GrantNo. AST-2034437 and a collaboration including Caltech, IPAC, the Weizmann Institute for Science, the Oskar Klein Center atStockholm University, the University of Maryland, Deutsches Elektronen-Synchrotron and Humboldt University, the TANGOConsortium of Taiwan, the University of Wisconsin at Milwaukee, Trinity College Dublin, Lawrence Livermore NationalLaboratories, and IN2P3, France. Operations are conducted by COO, IPAC, and UW. This publication makes use of data products from the Two Micron All Sky Survey, which is a joint project of the University of Massachusetts and the Infrared Processing and Analysis Center/California Institute of Technology, funded by the National Aeronautics and Space Administration and the National Science Foundation. he Pan-STARRS1 Surveys (PS1) and the PS1 public science archive have been made possible through contributions by the Institute for Astronomy, the University of Hawaii, the Pan-STARRS Project Office, the Max-Planck Society and its participating institutes, the Max Planck Institute for Astronomy, Heidelberg and the Max Planck Institute for Extraterrestrial Physics, Garching, The Johns Hopkins University, Durham University, the University of Edinburgh, the Queen's University Belfast, the Harvard-Smithsonian Center for Astrophysics, the Las Cumbres Observatory Global Telescope Network Incorporated, the National Central University of Taiwan, the Space Telescope Science Institute, the National Aeronautics and Space Administration under Grant No. NNX08AR22G issued through the Planetary Science Division of the NASA Science Mission Directorate, the National Science Foundation Grant No. AST-1238877, the University of Maryland, Eotvos Lorand University (ELTE), the Los Alamos National Laboratory, and the Gordon and Betty Moore Foundation. 

{\bf Online resources.} This research has made  use of the Simbad astronomical data base \citep{simbad} and the Vizier catalogue access tool \citep{vizier}, both operated at the Centre de Donn\'ees astronomiques de Strasbourg (CDS), and of \textsc{topcat} \citep{topcat}. The data analysis and figures were made with the \textsc{R} programming language \citep{R,data.table} and \textsc{rstudio} (\url{https://www.rstudio.com/}), or with \textsc{python} \citep[][\url{http://www.python.org}]{van1995python,numpy,scipy,astropy}. This research has made extensive use of the GaiaPortal catalogues access tool, Agenzia Spaziale Italiana (ASI) - Space Science Data Center (SSDC), Rome, Italy (\url{http://gaiaportal.ssdc.asi.it}); of the NASA Astrophysics Data System (\url{https://ui.adsabs.harvard.edu/}); of the arXiv preprint database (\url{https://arxiv.org/}). This research has made use of the SVO Filter Profile Service (\url{http://svo2.cab.inta-csic.es/theory/fps/}) supported from the Spanish MINECO through grant AYA2017-84089. This publication made extensive use of the collaborative cloud-based LaTeX editor Overleaf (\url{https://www.overleaf.com}).
\end{acknowledgements}


\bibliographystyle{aa} 
\bibliography{PhotSets} 

\end{document}